\documentclass{aastex631}

\usepackage{longtable}
\usepackage{multirow}
\usepackage{amsmath}
\usepackage{subfigure}

\usepackage{xcolor}
\definecolor{green1}{rgb}{0.2, 0.6, 0.1}
\definecolor{purp1}{rgb}{0.6, 0.1, 0.45}

\def\d{{\rm d}}
\newcommand{\rmd}{\mathrm{d}}
\newcommand{\dL}{$\delta \mathcal{L}^{\prime}$}
\newcommand{\dLc}{$\delta \mathcal{L}_c^{\prime}$}
\newcommand{\dLp}{$\delta \mathcal{L}_p^{\prime}$}
\newcommand{\Lc}{$\mathcal{L}_c^{\prime}$}
\newcommand{\Lp}{$\mathcal{L}_p^{\prime}$}
\newcommand{\Lwind}{$\mathcal{L}^{\prime}$}
\newcommand{\dH}{$\delta \mathcal{H}^{\prime}$}
\newcommand{\dHc}{$\delta \mathcal{H}_c^{\prime}$}
\newcommand{\dHp}{$\delta \mathcal{H}_p^{\prime}$}
\newcommand{\vzdl}{$v_z\,\delta \mathcal{L}_c^{\prime}$}
\newcommand{\vzdh}{$v_z\,\delta \mathcal{H}_c^{\prime}$}
\newcommand{\vzbz}{$v_z\,\left|B_z\right|$}

\newcommand{\kurtdL}{$\mathrm{Kurt}\left( \delta \mathcal{L}_c^{\prime} \right)$}

\newcommand{\kurtvzdl}{$\mathrm{Kurt}\left( v_z\,\delta \mathcal{L}_c^{\prime} \right)$}
\newcommand{\kurtvzdh}{$\mathrm{Kurt}\left( v_z\,\delta \mathcal{H}_c^{\prime} \right)$}
\newcommand{\kurtvzbz}{$\mathrm{Kurt}\left( v_z\,\left|B_z\right| \right)$}

\newcommand{\FAR}{65}
\newcommand{\NF}{79}
\newcommand{\TF}{144}

\newcommand{\deriv}[2]{\frac{\mathrm{d} #1}{\mathrm{d} #2}}
\def\ev{{\boldsymbol e}}
\def\uv{{\boldsymbol u}}
\def\rv{{\boldsymbol r}}
\def\xv{{\boldsymbol x}}
\def\yv{{\boldsymbol y}}
\def\d{\mathrm{d}}
\def\vv{{\boldsymbol v}}
\def\Bv{{\boldsymbol B}}

\received{December 5, 2024}
\submitjournal{ApJ}

\shorttitle{Dataset Preparation}
\shortauthors{Williams et al.}

\begin{document} 
    \correspondingauthor{Thomas Williams}
    \email{tomwilliamsphd@gmail.com}
    
   \title{Investigating the Efficacy of Topologically Derived Time-Series for Flare Forecasting.\\ I. Dataset Preparation}

   \author[0000-0002-2006-6096]{Thomas Williams}
   \affiliation{Department of Mathematical Sciences, Durham University, Durham, UK}

   \author[0000-0003-4015-5106]{Christopher B. Prior}
   \affiliation{Department of Mathematical Sciences, Durham University, Durham, UK}

   \author[0000-0003-2297-9312]{David MacTaggart}
   \affiliation{School of Mathematics \& Statistics, University of Glasgow, Glasgow, UK}

\begin{abstract}
The accurate forecasting of solar flares is considered a key goal within the solar physics and space weather communities. There is significant potential for flare prediction to be improved by incorporating topological fluxes of magnetogram datasets, without the need to invoke a three-dimensional magnetic field extrapolations. Topological quantities such as magnetic helicity and magnetic winding have shown significant potential towards this aim, and provide spatio-temporal information about the complexity of active region magnetic fields. This study develops time-series that are derived from the spatial fluxes of helicity and winding that show significant potential for solar flare prediction. It is demonstrated that time-series signals, which correlate with flare onset times, also exhibit clear spatial correlations with eruptive activity; establishing a potential causal relationship.  A significant database of helicity and winding fluxes and associated time series across \TF\ active regions is generated using SHARP data processed with the ARTop code that forms the basis of the time-series and spatial investigations conducted here. We find that a number of time-series in this dataset often exhibit extremal signals that occur $1$-$8$ hours before a flare. This, publicly available, living dataset will allow users to incorporate these data into their own flare prediction algorithms.
\end{abstract}
\keywords{solar flares -- space weather -- solar active region magnetic fields -- astronomy databases}

\section{Introduction}\label{sec:intro}
The forecasting of solar flares is one of the most important practical issues in solar physics. Solar flares have crucial impacts on how space weather affects critical infrastructure such as radio communication, GPS signalling, and power grids. A recent international collaboration performed a systematic comparison of existing flare forecast methods (as well as developing the statistical and methodological tools for this comparison) \citep{leka2019comparison,leka2019comparison3}. These methods are based on analysing photospheric magnetograms, which provide information on the structure of the Sun's magnetic field as it passes from the interior into the solar atmosphere. The efficacy of the results is briefly summarized by the following text from \citet{leka2019comparison}, 
``\textit{Regarding the results, generally speaking, no method works extraordinarily well; but we demonstrate that a fair number of methods consistently perform better than various no-skill measures, meaning that they do show definitive skill across more than one metric}''. Similarly, \citet{2021JSWSC} state, ``\textit{In spite of being one of the most intensive and systematic flare forecasting efforts to-date, FLARECAST has not managed to convincingly lift the barrier of stochasticity in solar flare occurrence and forecasting: solar flare prediction thus remains inherently probabilistic}''. In short, what these two studies reveal is that whilst significant progress has been made, there is still room for improvement. 

Magnetic topology, which can be quantified by the magnetic helicity as a measure of the flux-weighted entanglement of a magnetic field, has long been used as a diagnostic tool in the analysis of solar active regions e.g., \citet{pariat2006spatial, pariat2005photospheric, nindos2003magnetic,liu2012magnetic,knizhnik2015filament,priest2016evolution,pariat2017relative,zuccarello2018threshold,vemareddy2019very,liu2019formation,wang2018roles,jarolim2023probing,moraitis2024using}. The near perfect conservation of magnetic helicity in the corona means one can estimate the helicity content of active region fields from observational data (magnetogram data) via helicity fluxes at the photosphere \citep{pariat2005photospheric,park2010productivity,vemareddy2019very,park2021magnetic,artop1,korsos2020differences}.

Recently, two of the authors of this study demonstrated the efficacy of a related quantity, the magnetic winding: the entanglement of the field measured without flux weighting, which can also be estimated from observational magnetogram data \citep{prior2020magnetic}. It was shown to provide the first direct and unambiguous evidence of pre-existing twisted field structure emerging into the solar corona \citep{mactaggart21}. It was also shown that imbalances in current-carrying magnetic entanglement of developing active region fields presaged the onset of flaring in a set of active regions \citep{brenopaper,raphaldini2023deciphering} and, in addition, that significant localised variance in these signatures often trailed large X-class flares by $\approx 7$ hours, providing evidence of the potential for individual flare prediction. Finally in \citet{mnraspaper}, it was shown that spikes in the input of magnetic winding correlated consistently both temporally and spatially with the onset of coronal mass ejections (CME) in 28 of 30 active regions.

To promote the routine use of both magnetic helicity and winding flux estimation, Active Region Topology \citep[ARTop]{artop1} was released, an open source software which can calculate both helicity and winding fluxes as well as various combinations and decompositions of these quantities. The package provides time-series of the net values of these fluxes and also spatio-temporal time-series of their density distributions in the regions. The aim of this study is to establish, based on a much expanded dataset, the potential predictive efficacy of combinations of information produced by both the magnetic winding and helicity fluxes via ARTop with a particular focus on the spatial aspect of these distributions. In short, the ARTop package was aimed at generating meaningful spatiotemporal time-series from this data, and this study is aims to provide guidance as to how it might be used for flare prediction.

In \S\,\ref{sec:method} we present the methods adopted and dataset analysed in this study. In \S\,\ref{sec:results} a number of potential predictive quantities are investigated using three example active regions, whilst a more comprehensive dataset is explored in \S\,\ref{sec:results2}. Finally, our conclusions are presented in \S\,\ref{sec:conc}.

\section{Methodology}\label{sec:method}
Following a similar procedure to that outlined in \citet{mnraspaper}, we utilise two open-source codes, which are detailed below, to investigate how solar eruptions may be predicted. The first code calculates magnetic helicity and magnetic winding input rates at the photosphere and is written in C++ and python whilst the second code is an autonomous low-coronal CME detection code written in IDL. In the following subsections, the details of these two codes are summarised, while the full details on their respective methods can be found in \citet{artop1} and \citet{almanac}.

\subsection{Active Region Topology}\label{sec:artop}
Active Region Topology \citep[ARTop]{artop1} is an open-source tool for studying the input of topological quantities into solar active regions at the photospheric level. ARTop utilises vector magnetograms \citep{hoeksema14} from the Helioseismic and Magnetic Imager \citep[HMI]{hmipaper} aboard the Solar Dynamics Observatory (SDO) in the form of Space-Weather HMI Active Region Patches (SHARP) to create maps, time-series, and other metrics derived from input rates of magnetic helicity and magnetic winding fluxes (see \citealp{prior2020magnetic} for a detailed summary of their meaning and importance in Solar applications). The flux of magnetic helicity has long been considered an important quantity in the study of active regions and solar flares \citep{pevtsov2003helicity,park2008variation,vemareddy2021successive,korsos2022magnetic,liu2023changes}, the winding is a relativity novel quantity, which has been shown to have additional predictive efficacy \citep{prior2020magnetic,mactaggart21,brenopaper,mnraspaper} 

\subsubsection{Critical Quantities Provided by ARTop}\label{sec:CriticalQuants}
This study does not use all the quantities calculated by ARTop and so the following only provides a brief overview of the critical quantities that are directly used here. The fundamental geometrical quantity considered is the rotational motion of field line footpoints at the photospheric surface $P$. Let $\xv(t) = (x_1,x_2)$ and $\yv(t) = (y_1,y_2)$ represent the position vectors in $P$ of two field lines intersecting $P$ at a time $t$. The mutual angle $\Theta(\xv,\yv)$ of the two field line intersection points in $P$ is given by
\begin{equation}\label{theta}
\Theta(\xv,\yv) = \arctan\left(\frac{y_2-x_2}{y_1-x_1}\right).
\end{equation}
Its rate of rotation can be written as the in-plane motion of the footpoints $\uv(\yv) = (\d y_1/\d t,\d y_2/\d t)$, which is estimated using the DAVE4VM method \citep{schuck08}, and a vector $\rv=\yv-\xv$, joining the two points as: 
\begin{equation}\label{theta}
\deriv{\Theta(\xv,\yv)}{t} = \ev_z\cdot\frac{(\uv(\xv)-\uv(\yv))\times\rv}{|\rv|^2}\,.
\end{equation}
The magnetic winding input rate $d\mathcal{L}/\d t$, associated with a point $\xv$ in the photospheric plane $P$, is the average winding of that point with all other field line motions $\yv(t) \in P$:
\begin{equation}
\deriv{\mathcal{L}}{t}(\xv) = -\frac{1}{2\pi}\int_{P}\deriv{\Theta(\xv,\yv)}{t}\,\d^2y\,,
\end{equation}
and the field line helicity input rate $d\mathcal{H}/\d t$ is the winding weighted by the magnetic flux:
\begin{equation}
\deriv{\mathcal{H}}{t}(\xv) =-B_z(\xv)\frac{1}{2\pi}\int_{P}B_z(\yv)\deriv{\Theta(\xv,\yv)}{t}\,\d^2y\,.
\end{equation}
The minus signs represent the fact $P$ is the lower boundary of the domain in which the active region field exists (so the normal to P is opposite to the normal of that domain). As discussed in detail in \citet{prior2020magnetic}, this helicity input rate is the same as the relative helicity rates calculated in studies of photopsheric helicity fluxes \citep{park2008variation,vemareddy2021successive,korsos2022magnetic,liu2023changes}, we simply emphasise its geometric underpinning. ARTop also provides the spatially integrated winding $\d L/ \d t$ and helicity $\d H/ \d t$ inputs as
\begin{equation}
\deriv{L}{t} =\int_{P}\frac{\d\mathcal{L}}{\d t}(\xv)\,\d^2x\,, \quad \textrm{and} \quad \frac{\d H}{\d t} =\int_{P}\frac{\d\mathcal{H}}{\d t}(\xv)\,\d^2x,
\end{equation}
from which the time-integrated inputs can be calculated as
\begin{equation}
L(t) =\int_{0}^{t}\frac{\d L}{\d t} \d t,\quad \textrm{and} \quad H(t) =\int_{0}^{t}\frac{\d H}{\d t} \d t.
\end{equation}
Approximate helicity conservation implies that $H(t)$ should be a good estimate of the amount of helicity in the active region field above the photosphere \citep{berger1984rigorous}, and this manuscript focuses on how crucial this estimation can be for flare prediction, a fact which has previously been observed in other studies \citep{park2008variation,labonte2007survey,jarolim2023probing,garland24}. It was shown in \citet{prior2020magnetic,mactaggart21,brenopaper,mnraspaper} that the winding provides distinct and complementary information to the helicity, as the flux weighting in the helicity means it is dominated by magnetic field with a strong vertical component and twisted nature (i.e. the main poles) whilst the winding is largely dominated by strong (dominantly) transversal field near the polarity inversion line (i.e. the top of sheared arcades or bald patch field).

Field line velocities, can be estimated from magnetogram data by assuming ideal motion \citep{pariat2005photospheric}. Under this assumption, the motion $\uv(\xv)$ of the point of intersection of a field line and the photosphere can be written in terms of the field $\mathbf{B}$ and the plasma velocity $\vv(\xv)$ (decomposed into out-of-plane $v_z$ and in-plane $\vv_{\|}$ components) as:
\begin{equation}\label{eq:veleq}
\uv(\xv) = \vv_{\|}(\xv) - \frac{v_z(\xv)}{B_z(\xv)}\Bv_{\|}(\xv) = \uv_b + \uv_e,
\end{equation}
where $\uv_b$ represents the plasma moving the field line ideally in-plane (braiding if adding to the winding) and the second term $\uv_e$ represents the emergence/submergence of field. The velocity has implicit ${\bf B}$-dependence as it is determined by DAVE4VM using the magnetic field data. We highlight that this inversion has a crucial parameter associated with it; the magnetic field must be smoothed for the least square matrix used in the inversion of the DAVE4VM method to be well defined, we term it VS (velocity smoothing) in ARTop. VS is the width (number of pixels) of the window surrounding the point of interest used to locally average the field. The velocity and helicity values can be significantly affected by this choice \citep{schuck08,bi2018survey,artop1}. The work presented here develops quantities whose predictive efficacy are as independent of this choice as possible.

The final critical quantities to introduce are those on which our metrics are based. It is possible to decompose the field into a potential part and a ``current-carrying'' part, \textit{i.e.} the Helmholtz decomposition:
\begin{equation}\label{decomp}
    \Bv(t) = \Bv_p(t) + \Bv_c(t). 
\end{equation}
The potential part is uniquely determined by the $B_z$ distribution on the phostopheric boundary \citep[see a discussion on the method used]{brenopaper,artop1}, which then gives $\Bv_c(t)$ from the observed in-plane components. Using these two fields one can calculate current-carrying helicity and winding fluxes, $\d\mathcal{H}_c/\d t$ and $\d\mathcal{L}_c/\d t$, and potential fluxes, $\d\mathcal{H}_p/\d t$ and $\d\mathcal{L}_p/\d t$. Then, since we expect flaring to occur when there is a (local) imbalance towards current-carrying topology, ARTop calculates the following $\delta$ quantities
\begin{equation}\label{eq:dL}
\delta L(T) = \int_0^T\int_P\left(\left|\deriv{\mathcal{L}_c}{t}\right| - \left|\deriv{\mathcal{L}_p}{t}\right|\right)\,\rmd^2y\,\rmd t,
\end{equation}
and
\begin{equation}\label{eq:dH}
   \delta H(T) = \int_0^T\int_P\left(\left|\deriv{\mathcal{H}_c}{t}\right| - \left|\deriv{\mathcal{H}_p}{t}\right|\right)\,\rmd^2y\,\rmd t,
\end{equation}
which is positive if there is an imbalance towards current-carrying winding/helicity flux. In \citet{brenopaper,artop1} it was shown the $\delta$ fluxes were only affected a small amount by the choice of velocity smoothing, VS, by contrast it had quite a significant effect on the rates $\rmd\mathcal{L}_c/\rmd t$, $\rmd\mathcal{H}_c/\rmd t$, $\rmd\mathcal{L}_p/\rmd t$ and $\rmd\mathcal{H}_p/\rmd t$. This study provides further evidence that quantities based on the $\delta$ quantities are far more consistent with regards to flare prediction signals than the rates $\rmd\mathcal{H}/\rmd t$ and $\rmd\mathcal{L}/\rmd t$ themselves. In what follows, for the sake of notational brevity, we denote all rates with a dash \textit{i.e.} $\mathcal{H}' =\rmd\mathcal{H}/\rmd t$ or $\delta L'=\rmd\delta L/\rmd t$.

\subsection{ALMANAC}\label{sec:almanac}
As with \citet{mnraspaper}, we also employ the Automated Detection of CoronaL MAss Ejecta origiNs for Space Weather AppliCations (ALMANAC) code \citep{almanac} when focusing on specific events in time-series data to identify potential CMEs. ALMANAC, unlike many widely adopted CME detection methods does not rely upon coronagraph data, but instead utilises data from the Atmospheric Imaging Assembly \citep[AIA]{lemen12}. The main advantage of ALMANAC is that it does not require geometrical fitting to approximate the CME source location in the low solar corona. Subsequently, the code does not inherently have large uncertainties due to projection effects caused by fitting a simple ``wire-frame'' of a three-dimensional object mapped in two dimensions. As such, ALMANAC provides a reliable low-coronal CME origin that is obtained independently of any helicity/winding signatures from the earlier phases of an eruption.

To detect potential Earth-directed CMEs, ALMANAC first crops the map size to eliminate off-limb contributions and standardises the intensity across an 8-hour image sequence by thresholding intensities and normalising the data values. It is then smoothed through convolution and subtracted from the normalised data to create a high-bandpass and time-filtered image sequence. Each time step of the time-filtered data is then divided by the median of the absolute values of the unfiltered data to eliminate contribution from ``static'' structures such as active regions. The method employs a series of Boolean masks to isolate connected clusters of pixels associated with a potential eruption, and spatio-temporal smoothing of these masks helps avoid the segmentation of regions. The first time step in which a region of sufficient size and duration is identified is used as the CME onset time, whilst the centre of mass for the masked pixels at that time provides the central location for the CME. Full details of the method can be found in \citet{almanac}.

\subsection{Kurtosis Analysis}\label{sec:KurtAnal}
One additional quantity we consider in this study is the Kurtosis of the ARTop derived time series. Kurtosis is a statistical quantity used to characterise the relative ``fatness'' of the tails for a probability distribution relative to the mean of the distribution. For example, a normal distribution has a bell curve shape when plotted as a histogram, with the majority of the data-points residing within three standard deviations of its mean. As such, a normal distribution has a kurtosis of 3 and would be called a mesokurtic distribution. A mesokurtic distribution is often considered to depict a moderate level of risk. Following this, the excess kurtosis is the difference between the kurtosis of the measured distribution in relation to a normal distribution. In this instance, the excess kurtosis of a mesokurtic distribution is 0.

\begin{figure}
\centerline{\includegraphics[width=0.5\textwidth]{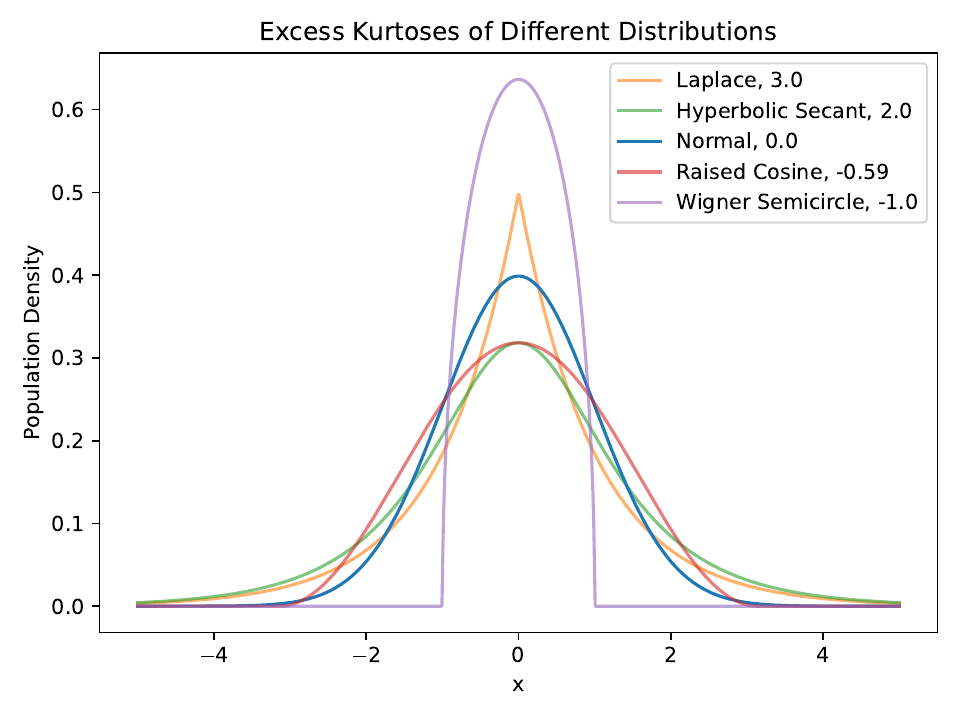}}
\caption{Examples of kurtoses (legend) for different distributions with $\mu=0,\,\,\sigma=1$.}
\label{fig:kurt_example}
\end{figure}

If the excess kurtosis $<0$, the distribution is considered to be platykurtic and the distribution will have shorter tails, or fewer outliers relative to the mean than a normal distribution. In this instance, the distribution may be considered more stable than a state of mesokurticity because extreme curve movements have rarely occurred in the past. On the other hand, lepotkurtic distributions have excess kurtoses $>0$ and subsequently have larger tails than a normal distribution. These appear as slim-peaked distributions due to a large number of outliers stretching the horizontal axis (Figure\,\ref{fig:kurt_example}). In contrast to a platykurtic distribution, a leptokurtic distribution depicts a high level of risk due to a more extreme ratio of outliers compared to values near the mean. Rapid increases in the excess kurtosis has been shown to be an important precursor to system bifurcations (or critical transitions) in a number of fields - for example, in ecological systems \citep{dakos2019ecosystem}, climate \citep{boers2018early}, the economy \citep{sevim2014developing}, and medicine \citep{chen2012detecting}.

\subsection{Dataset}\label{sec:data}
\begin{figure}
\centerline{\includegraphics[width=0.5\textwidth]{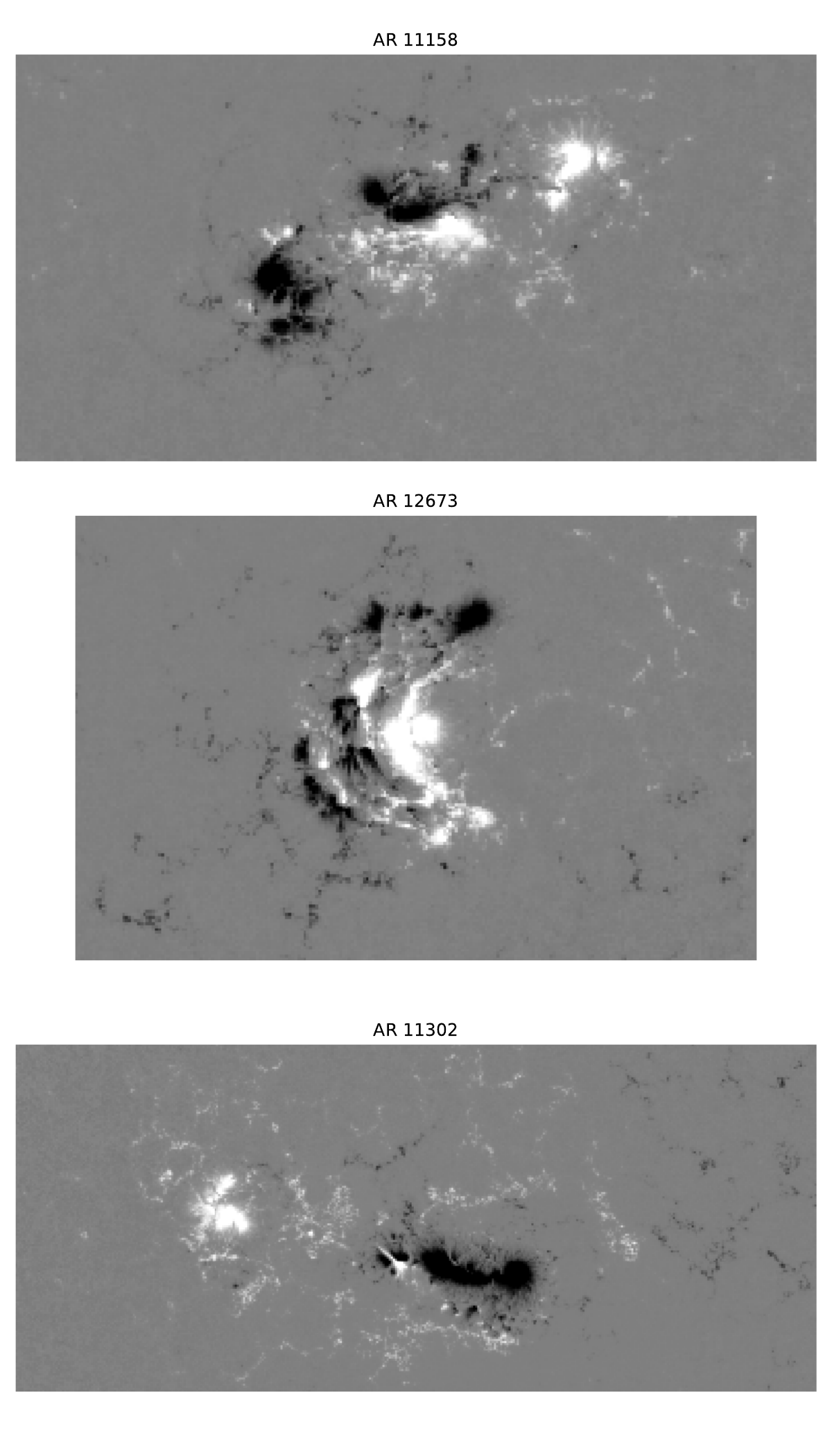}}
\caption{SHARP vector magnetograms from HMI for the three active regions discussed in detail in section\,\ref{sec:results}.}
\label{fig:bzmaps}
\end{figure}

In \S\,\ref{sec:results} we initially analyse a subset of three active regions (Table\,\ref{tab:sharps}). These examples are used to highlight some critical aspects underlying the construction of the new metrics presented in this work, before investigating some of the conclusions from this set on a much larger sample of \TF\ SHARP regions (see Appendix for details). From these examples, two are what may be considered `classic' examples, such as AR\,11158 and AR\,12673, for their predictable behaviour and propensity for large eruptions. The three regions we highlight to demonstrate new quantities that could potentially aid flare prediction are all distinctly different, as can be seen from their HMI vector magnetograms in Figure\,\ref{fig:bzmaps}.

AR\,11158 begins as a pair of aligned bipoles. As it evolves, the inner poles of each bipole are seen to interact and rotate forming a more complex morphology. The topological properties of this region have been discussed in numerous studies \citep{tziotziou2013interpreting,lumme2019probing,thalmann2019magnetic,korsos2022magnetic,jarolim2023probing}, whilst \citet{brenopaper} provides a detailed comparison of the ARTop time-series analysis of this region in relation to those studies. These precise dynamics are not directly relevant to the results presented here, suffice to say all studies note significant helicity injection due to the the mutual rotation of the inner poles and their subsequent interaction.

For the first 120\,hours of AR\,12673 observations, it is a single positive polarity pole, which undergoes a sudden emergence of positive and negative magnetic field, with the negative field wrapping around the active region, forming multiple strong polarity inversion lines that expand rapidly as new emergent field `pushes' existing field into opposite polarity structures. As with AR\,11158, the topological properties of this region have been discussed in numerous studies \citep{moraitis2019magnetic,vemareddy2019very,thalmann2019magnetic,price19,kusano2020physics,korsos2022magnetic}, and a detailed comparison of the ARTop time-series analysis of this region by comparison to those studies is conducted in \citet{brenopaper}. The analysis of a large spike in the winding flux time-series just prior to the two-large X-class flares which were shown to spatially coincide with a flux rope in a non-linear force-free field extrapolation of this region performed by \citet{kusano2020physics} is of particular interest. It was shown that the spike coincided with a downflow (negative velocity plasma flow) in a region with a highly concentrated shear.

As for AR\,11302, this is largely bipolar with some parasitic negative polarity field in the centre of the region that encompasses a portion of the positive polarity field. As the region evolves, the positive polarity disperses into a diffuse field, with the negative polarity pole bifurcating into several smaller `poles' before it is swept beyond HMI's field-of-view. 

These three regions provide a varied basis upon the conditions for flaring to occur, and thus serve as the initial testing for the feasibility of quantities to be flare predictors. The three active regions (Table\,\ref{tab:sharps}) are then analysed to quantify the flaring in relation to the parameters discussed throughout \S\,\ref{sec:results}. In \S\,\ref{sec:results2}, we further explore these parameters on a more complete dataset that is detailed in the Appendix.

\begin{table}
\caption{SHARP regions investigated with flare information provided by the Heliophysics Event Knowledgebase \citep[\textit{HEK}]{hekpaper}.}
\label{tab:sharps}                  
\centering                          
\begin{tabular}{c c c c c c c}       
\hline                       
NOAA Active & SHARP Number & Largest Flare & \# X--class & \# M--class & \# C--class & First Observation \\
Region & & & Flares & Flares & Flares & Time (UTC) \\
\hline                             
11158 & 377 & X2.2 & 1 & 3 & 24 & 2011/02/10 22:58:11 \\
11302 & 892 & X1.9 & 2 & 15 & 30 & 2011/09/21 12:34:20 \\
12673 & 7115 & X9.3 & 3 & 12 & 20 & 2017/08/28 08:58:43 \\ 
\hline                                   
\end{tabular}
\end{table}

\section{Determination of Meaningful Quantities}\label{sec:results}
\subsection{Velocity Smoothing and Downsampling}\label{sec:dataset}

\begin{figure*}
\centerline{\includegraphics[width=\textwidth]{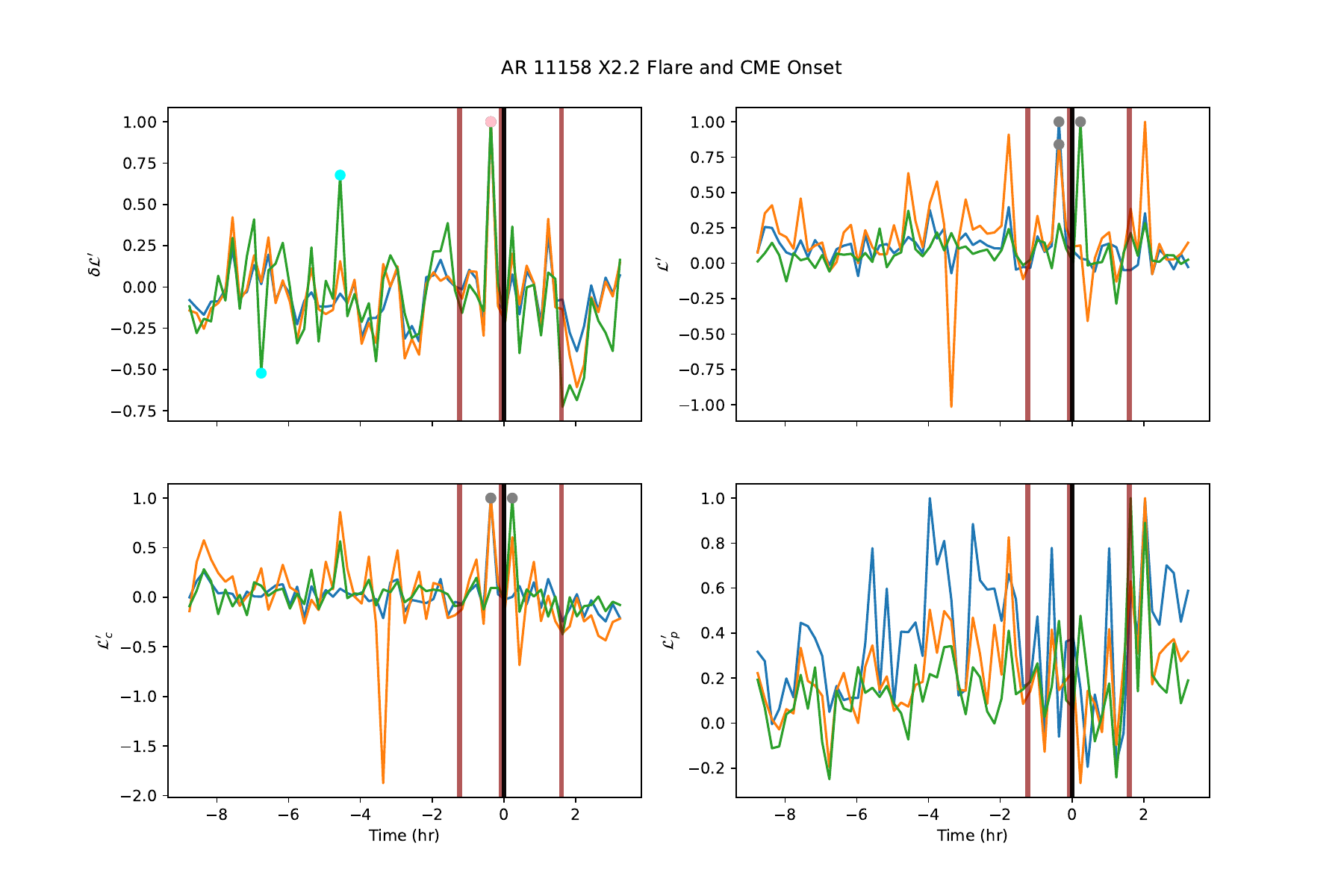}}
\vspace*{-3ex}
\begin{center}
\textbf{Blue: } $VS = 20,\,\,D=1$\quad\textbf{Orange: } $VS=20,\,\,D=3$\quad\textbf{Green:} $VS = 12,\,\,D=3$
\end{center}
\caption{Panels indicate the time-series evolution for topological quantities before and after the X2.2 flare and sympathetic CME of AR 11158 for various velocity smoothing and downsampling values. All plots are normalised with respect to their largest value. The X2.2 flare start time has been taken as Time = 0, which is indicated in \textit{black} with the CME onset times reported by ALMANAC shown in \textit{red}. The \textit{grey}, \textit{pink}, and \textit{cyan} circles denote the peaks discussed in the manuscript where spikes are (or not) shifted due to the choice of $VS$.}
\label{fig:velcomp}
\end{figure*}

In this subsection, the focus is on differences detected when utilising different parameter values within the ARTop code. To highlight these differences we focus on the large eruption associated with NOAA AR 11158, which led to an X2.2 solar flare and sympathetic CME four minutes later that was detected in SOHO/LASCO data \citep{mnraspaper}, as indicated in Figure\,\ref{fig:velcomp} on various ARTop derived time-series. In addition to this large eruption, ALMANAC also detected two smaller CMEs that occurred 1.25\,hours prior to, and 1.67\,hours post the X2.2 eruption, respectively. This is one example, from a set of 30 events in \citep{mnraspaper}, of a CME event which was shown to be presaged by a significant spike in the rate \Lwind. These spikes were also correlated to key magnetic structures in extrapolations of the field, hence they are shown to be physically meaningful.

As is discussed in \S\,\ref{sec:CriticalQuants}, ARTop utilises the DAVE4VM method to calculate velocity from noisy magnetogram data which employs a velocity smoothing ($VS$). The value of $VS$ represents a padding of typically between 11-20 pixels \citep{mactaggart21, schuck08} surrounding the pixel for which a velocity is being determined. In \citet[Figure 5]{artop1}, an example is shown for values of $VS = 12$ and $20$ pixels that indicates no significant difference in the general behaviour of the time-series for the topological quantities calculated. In \citet{brenopaper} it was found however, that some quantities derived from these calculations can show significant differences depending on this choice. The finding from \citet{brenopaper} is that the quantity \dL\ is relatively consistent to this value, a finding we elaborate on here. In Figure\,\ref{fig:velcomp}, we highlight differences in the time at which a peak is seen for $VS = 12$ (\textit{green}) and $20$ (\textit{orange}) pixels in the magnetic winding rate over both of its decompositions into current-carrying and potential components. The most notable difference here is that the examples of \Lwind\ (and \Lc ) utilising $VS=20$, which display significant peaks prior to the eruptions that are not captured until post-eruption for $VS=12$ pixels (\textit{grey circles} in Figure\,\ref{fig:velcomp}). As with \citet{artop1}, Figure\,\ref{fig:velcomp} highlights that the downsampling factor, $D$ applied to increase the processing speed of ARTop has negligible effects on these quantities (\textit{blue}, $D=1$; \textit{orange}, $D=3$).

In Figure\,\ref{fig:velcomp}, the input rates of \dL, \Lwind, \Lc, and \Lp\ for various choices of the parameters $VS$ and $D$ are presented. Crucially, this indicates that \dL\ is not particularly sensitive to the choice of $VS$ and $D$ made during data processing with ARTop as all three time-series exhibit temporally coincident peaks that are immediately before the X2.2 flare and CME (\textit{pink circle} in Figure\,\ref{fig:velcomp}). Unlike the other magnetic winding parameters shown, all time-series typically peak and trough at the same time often with similar magnitudes. There are, however, instances such as Time $\approx -6.5 \text{\,hours and } \approx -4.5$\,hours where the peak (trough) in the data processed with $VS = 12$ (\textit{cyan circles} in Figure\,\ref{fig:velcomp}) is larger than the peaks (troughs) seen for the two $VS = 20$ calculations. 

Since the choice of the parameter $VS$ is somewhat arbitrary (and has taken numerous different values in the literature) we make the decision to base the metrics we develop in what follows on the parameter $\delta L$, which is the least sensitive to this choice. This decision was validated by verifying that the database of physically meaningful spikes in \citet{mnraspaper} also exhibits spikes in \dL\ at the same times found in this study.

\subsubsection{Decomposing \dL\ and \dH}\label{sec:decomposing}
The main implication of the result shown in Figure\,\ref{fig:velcomp} is that the decomposition of magnetic winding (and magnetic helicity) into components for current-carrying and potential topology are sensitive to the parameter $VS$. Given the lack of sensitivity of \dL\ to these parameters, we instead split \dL\ and \dH\ into positive and negative components whereby positive (negative) would indicate that the dominant input of topology at the photospheric level is current-carrying (potential) topology. This is achieved by turning equations\,(\ref{eq:dL}) and (\ref{eq:dH}) into conditional expressions, such that:
\begin{equation}\label{eq:dLc_dHc}
   \delta\mathcal{L}_c^{\prime} =
    \begin{cases}
       \delta\mathcal{L}^{\prime}, & \text{if } \delta\mathcal{L}^{\prime} > 0 \\
        0,                                       & \text{otherwise}
\end{cases}
\quad\text{,}\quad
    \delta\mathcal{H}_c^{\prime} =
    \begin{cases}
       \delta\mathcal{H}^{\prime}, & \text{if } \delta\mathcal{H}^{\prime} > 0 \\
        0,                                       & \text{otherwise}
\end{cases}
,
\end{equation}
and
\begin{equation}\label{eq:dLp_dHp}
    \delta\mathcal{L}_p^{\prime} =
    \begin{cases}
        \delta\mathcal{L}^{\prime}, & \text{if } \delta\mathcal{L}^{\prime} < 0 \\
        0,                                       & \text{otherwise}
\end{cases}
\quad\text{,}\quad
    \delta\mathcal{H}_p^{\prime} =
    \begin{cases}
        \delta\mathcal{H}^{\prime}, & \text{if } \delta\mathcal{H}^{\prime} < 0 \\
        0,                                       & \text{otherwise}
\end{cases}
.
\end{equation}

Thus, these decompositions can be used to focus solely on the net input of  either current-carrying dominant topology (\dLc, \dHc) or potential-field dominant topology (\dLp, \dHp) at the photosphere (rather than allowing the two to cancel as in the series $\delta\mathcal{L}$). It has previously been demonstrated that it is the current-carrying component of \Lwind\ that is likely responsible for disrupting an existing magnetic field in the solar atmosphere that leads to an eruption (as in \citealp{pariat2017relative,brenopaper}) and so we will consider time series of based upon \dLc\ and \dHc\ here.

\begin{figure*}
\centerline{\includegraphics[width=0.7\textwidth]{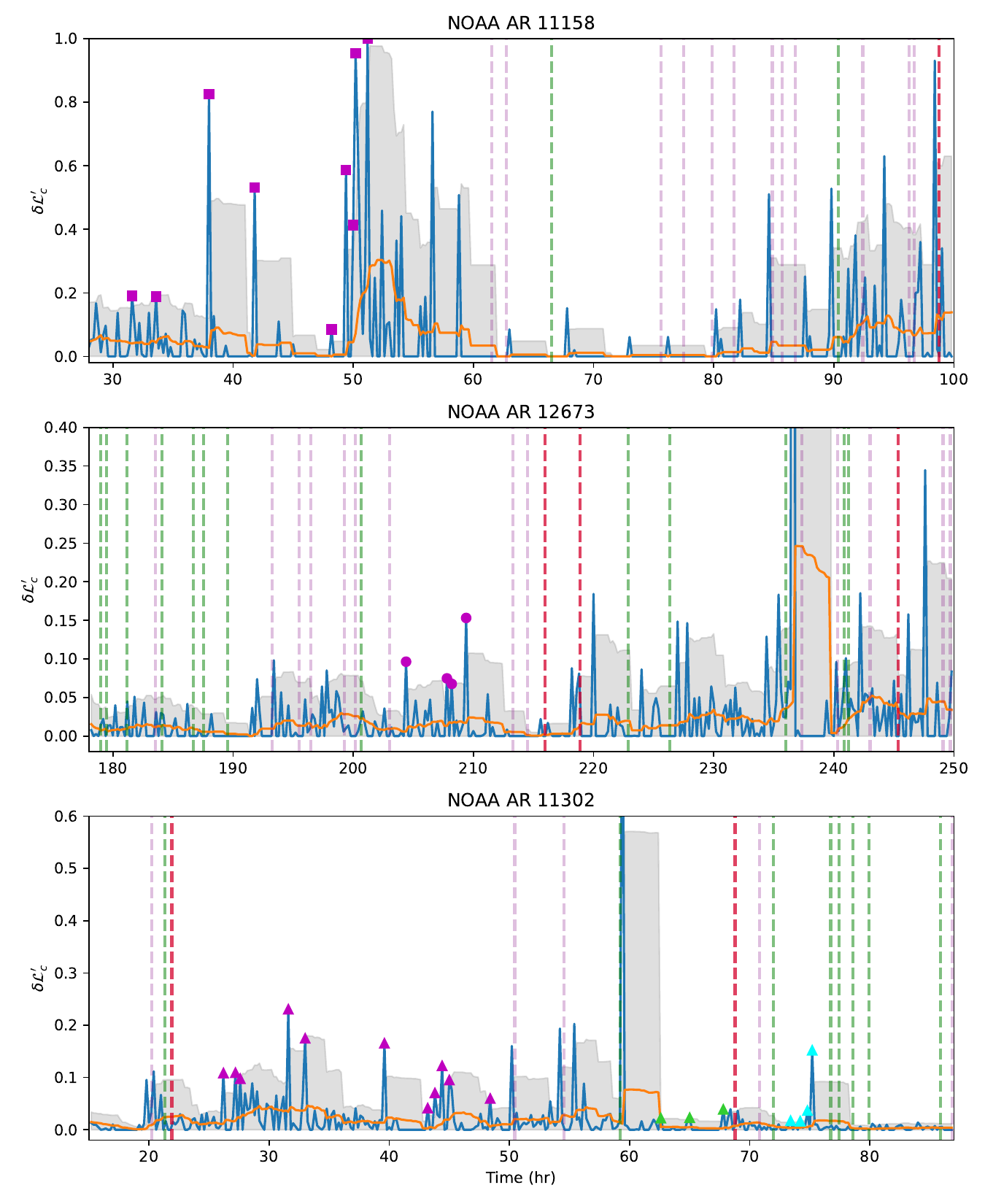}}
\caption{Time-series plots for \dLc\ (\textit{blue}) with a running mean of 3 hours (\textit{orange}) and $2\sigma$ envelope (\textit{grey}) for three SHARP regions preceding M-- (\textit{dashed green}) and X--class (\textit{dashed red}) solar flares. For context, C--class flares are also shown (\textit{dashed pink}). Each plot is normalised with respect to the maximum spike in the time-series.}
\label{fig:dl}
\end{figure*}

\begin{figure*}
\centerline{\includegraphics[width=0.7\textwidth]{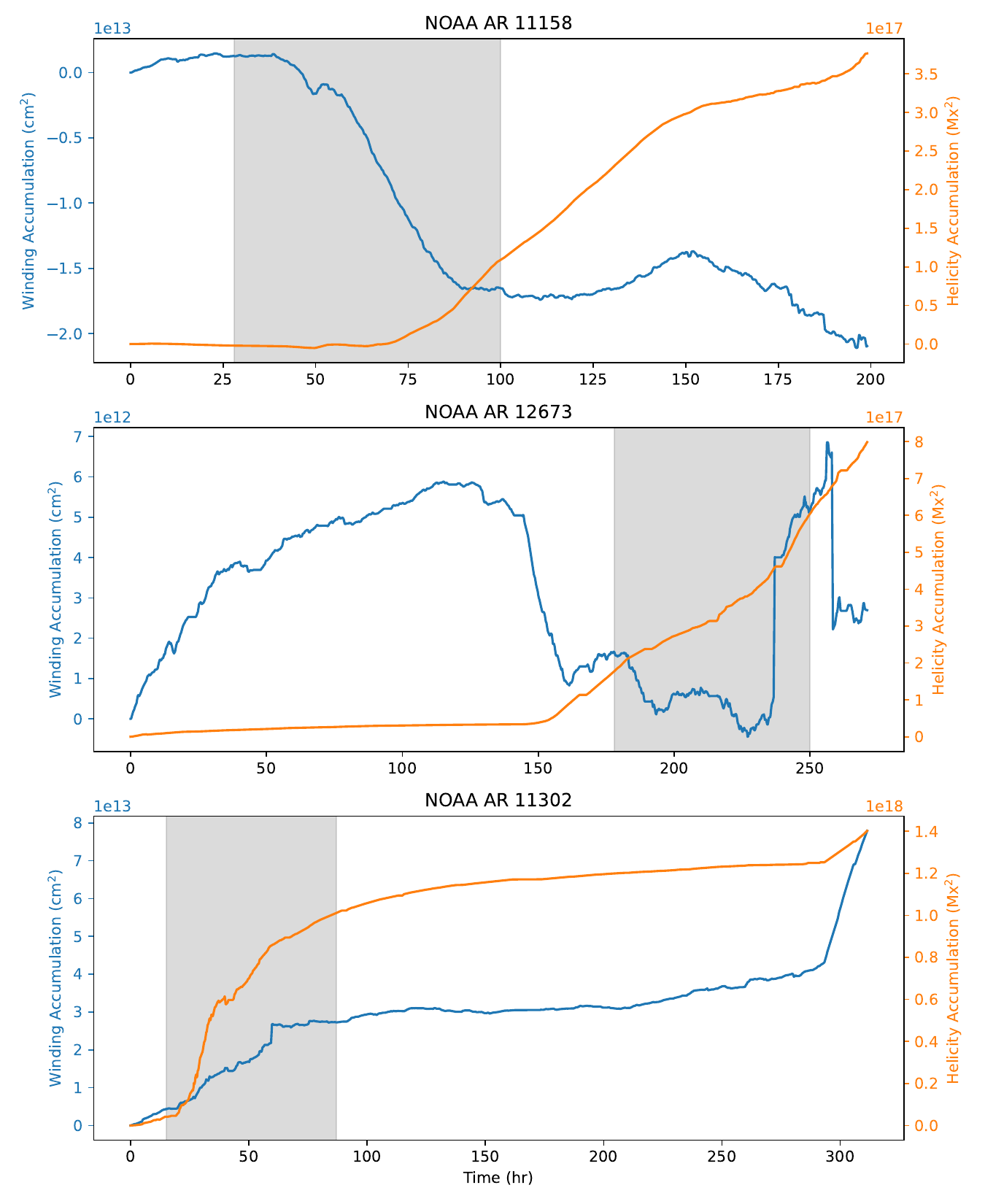}}
\caption{The time-integrated quantities \dL\ (\textit{blue}) and \dH\ (\textit{orange}) calculated for active regions NOAAs 11158, 11302, and 12673. The shaded gray regions indicate the 72-hour observation windows shown in Figure\,\ref{fig:dl} for each active region. The large increases seen in the winding accumulation, and to a lesser extent helicity accumulation, towards the end of the observed time for AR\,11302 and AR\,12673 are caused by projection effects in the SHARP data due to the active regions exceeding longitudes of 60$\,^{\circ}$.}
\label{fig:accum}
\end{figure*}

In Figure\,\ref{fig:dl}, \dLc\ time-series plots are shown for three example SHARP regions over 72-hour periods that include a range of C-- to X--class flares, which are used to highlight some of the properties of this quantity with regards to the potential for flare prediction. A 3-hour running mean and corresponding $2\sigma$ envelope are calculated for each series, which are used to determine significant spikes/peaks in the time-series that could lead to an eruption \citep{brenopaper,mnraspaper}. Here, a spike, or an extremal event, is defined as a point in time when the time-series in question exhibits a value outside the $2\sigma$ envelope\footnote{It is worth noting that the selection of the envelope is somewhat arbitrary, with \citet{brenopaper} utilising $3\sigma$ whilst \citet{mnraspaper} demonstrates that $2\sigma$ is a reliable threshold for a sample of 30 different active regions.}. 

Focusing first on NOAA AR\,11158, we see there are nine spikes during Time $ = 30-55$\,hours (\textit{magenta squares}) that are not followed by flaring within $6-12$\,hours of a spike, whilst the majority of spikes indicated beyond this time either precede large M-- and X--class flares or they are followed by multiple smaller C--class eruptions within $6-12$\,hours. This suggests one needs additional information to determine if such spikes in a time-series are (potentially) meaningful. To this end, analysis of the time-integrated quantities $\delta \mathcal{L}$ and $\delta \mathcal{H}$ (Figure\,\ref{fig:accum}) reveals that the spikes are only followed by flares (within a $6-12$\,hour period) after there has been a significant cumulative input of current-carrying helicity. One possible interpretation of this observation is that the spikes in \dLc\ seen between Time = $30$-$55$ hours (Figure\,\ref{fig:dl}) are not meaningful as there is insufficient complex overlaying magnetic field structure to disrupt and initiate flaring. We investigate this particular period in more detail later when assessing spatial correlations between events.

This observation is supported by the activity in the case of AR\,12673, we highlight a period between $180$ and $250$ hours where, we see in Figure\,\ref{fig:accum} that there has already been significant net current-carrying helicity input $\delta H$. In this case almost all spikes can be correlated with a flare given a window of approximately 6 hours. We now focus on one specific set of signals within this window. First between $\approx 203$\,hours and $\approx 210$\,hours, where there are relatively large spikes (\textit{magenta circles}) in the time-series that exceed their $2\sigma$ envelope (but no flares until $\approx 213$\,hours to $\approx 220$\,hours), whence there are two C--class and two X--class flares; the largest apparent gap in this series between a set of spikes and a flare. This contrasts significantly to the period before, $180$-$200$ hours, where there is a relatively steady occurrence of both smaller spikes and flares. These highlight the possibility of a pause in activity during which significant build-up of topology leads to the two significant X-class flares. We investigate this period in more detail later in the study when we focus on spatio-temporal correlations.

For AR\,11302 (bottom panel of  Figure\,\ref{fig:dl}), the first X--class flare emitted occurs during the `build-up' phase of the active region (around $21$ hours), which is denoted by the fact that the cumulative winding (Figure\,\ref{fig:accum}) continually increases until Time $\approx 70$\, hours, after which point the gradient flattens and a `steady-state' is reached (such as is discussed in \citealp{brenopaper}). Initially, this event appears somewhat unusual as our rough assumption from the previous examples is that active regions require time to develop complex topology before larger eruptions occur. However, ARTop does not observe AR\,11302 from its initial emergence as the active region rotates into view already partly developed, meaning the time-integrated inputs are not truly reflective of the total amount of helicity/winding injected into the region. This provides an example for one of the potential issues with building a predictive flaring model from photospheric topology inputs alone. That is, if an active region comes into view by rotation whereby a significant portion of the active region emergence phase has already occurred, then some additional function or method must be used to approximate the complex topology that has already been inputted into the solar atmosphere. One such possible method would be to use non-linear force-free field extrapolations to estimate the topology in the region as early as possible as in \citet{jarolim2023probing}. From the standpoint of eventually developing a live predictive flare model the fact there is a lull in flaring (Time $\approx 20-50$\,hours) is also of interest. It seems the first X--class flare may have resulted in a significant decrease of complex topology above the photosphere, and so one might infer that additional complexity must be rebuilt again prior to additional eruptions occurring. Thus, when building a live predictive flare model based on helicity/winding calculations, one would need to account for the decrease in complex magnetic field within the active region after an eruptive event, as this is something that cannot be accounted for from photospheric calculations alone. However, it might be possible to estimate/account for the loss of complex field from the size of the flare alone with the adoption of additional quantities explored later in this manuscript and machine learning.
 
 Later in the AR\,11302 time-series, when significant current-carrying helicity has been injected into the region, we see another interesting feature of these time-series, there are some relatively small amplitude spikes (\textit{green triangles}) which precede relatively strong M-- and X--class flares. By contrast, we later see three `small' and one `large' spike (\textit{cyan triangles}) which are seen to precede a burst of four M--class flares in an $\approx 4$\,hour period. There does not seem to be a clear correlation between the magnitude of a spike and the magnitude of the flare following it. We do note, however, in the case of the X-class flare, prior to this event at $\approx 69$ hours there is a significant input of $\delta L'$ about ten hours prior. This could potentially indicate that if information is included from a larger time window there may have been some indication of the potential for a large flare. We shall explore this event in more detail later in the study. 

\subsubsection{Discussion}
From these example calculations we have seen a number of interesting properties which will be investigated further in the rest of the study.
\begin{itemize}
\item{There are very often extremal spikes in the rates of current-carrying-dominant topology input, \dLc\ and \dHc, in a period up to 12 hours prior to flaring, with 69.7\% of \dLc\ spikes, and 65.3\% of \dHc\ spikes preceding flares in the three example regions.}
\item{These spikes correlate better to flaring activity when there has been a significant build up of (non potential field) helicity built up in the region. For example, when helicity accumulation exceeds $1\times10^{19}\,\mathrm{Mx}^2$, spikes preceding flares increase to 78.5\% and 76.9\% for \dLc\ and \dHc, respectively.}
\item{If the initial emergence phase of the region is not captured in the data it may be necessary to invoke additional information to estimate the helicity in the field when it emerges into view.}
\item{The magnitude of these spikes alone do not seem to correlate well to the size of the flare which lies in its 6-hour post-spike window. Subsequently, more information is required to make specific predictions on flare magnitude.}
\end{itemize}

\subsection{The input and loss of magnetic topology}\label{sec:combining}
\begin{figure*}
\centering
\subfigure{\includegraphics[width=0.8\textwidth]{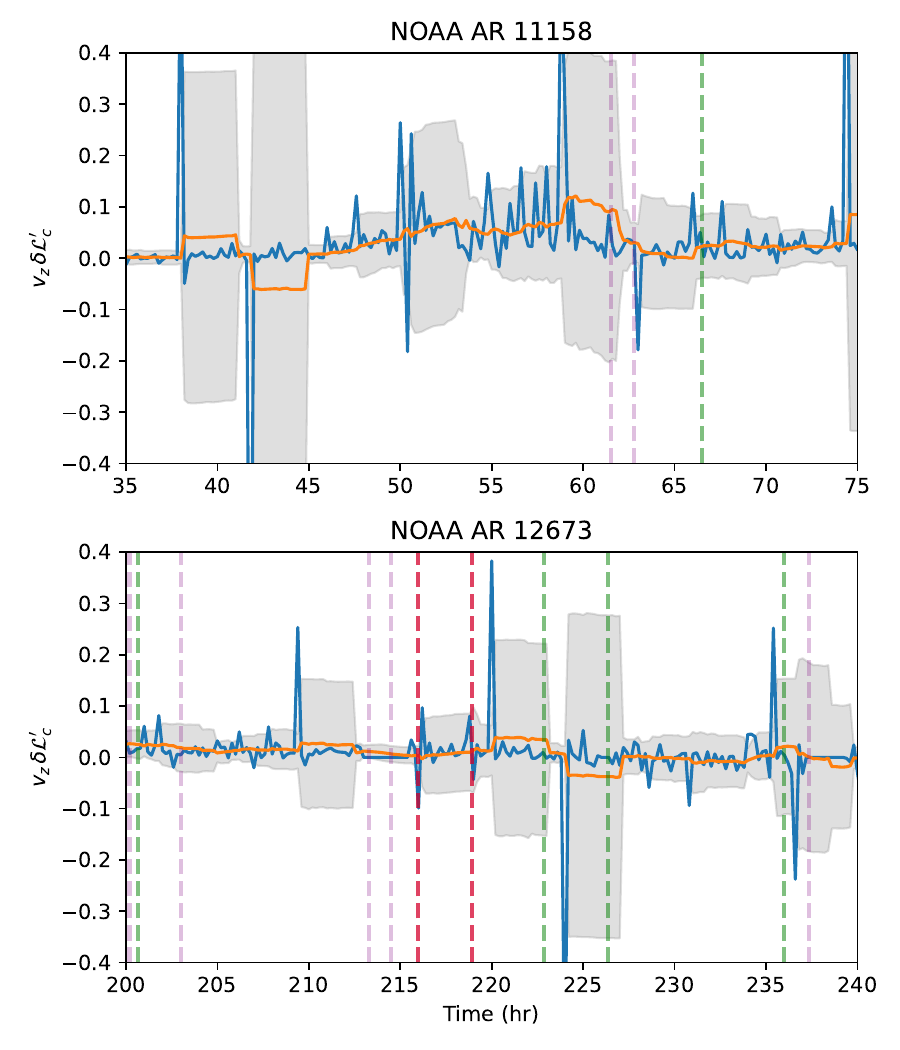}}
\caption{The time-series plots for the combined quantity, \vzdl\ focusing upon the period leading up to the onset of flaring for AR\,11158 and the X2.2 and X9.3 flares in AR\,12673. The 3-hour running means (\textit{orange}) and $2\sigma$ envelope (\textit{gray}) are also shown along with the times of C- (\textit{dashed pink}), M- (\textit{dashed green}), and X-class (\textit{dashed red}) flares.}
\label{fig:vzdl}
\end{figure*}
\begin{figure*}
\centering
\subfigure{\includegraphics[width=\textwidth]{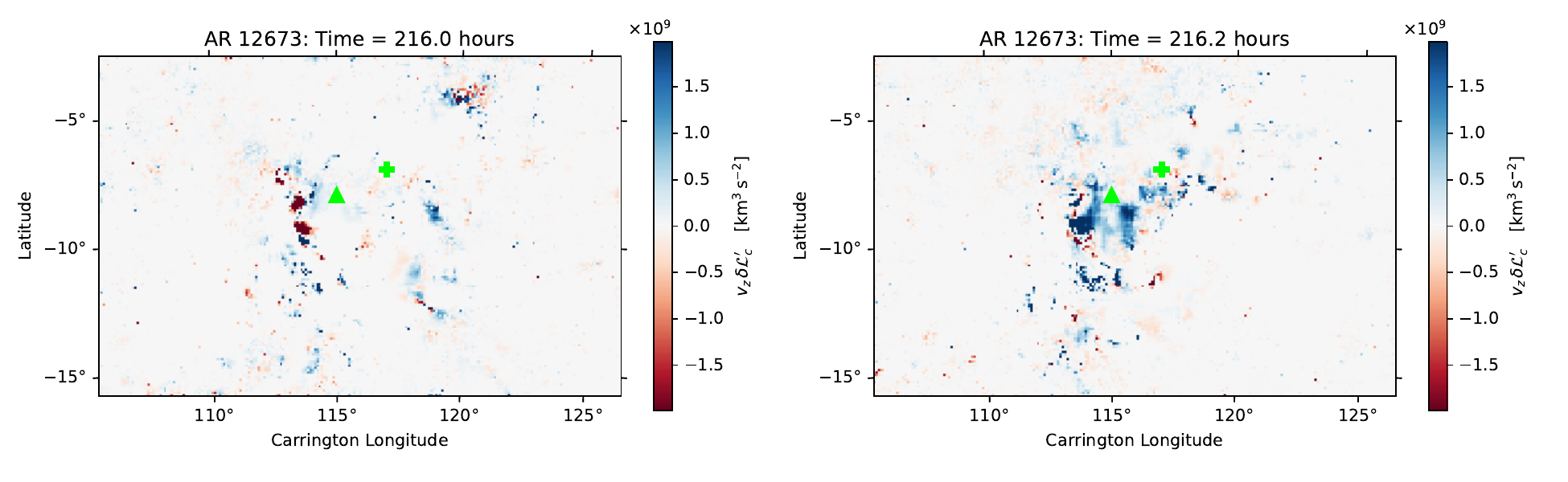}}
\caption{AR\,12673 \vzdl\ maps that correspond to the submergence and re-emergence of topology seen for the first X-class flare (\textit{green plus}) at Time = 216\,hours (first \textit{dashed red} line in Figure\,\ref{fig:vzdl}). The position of the second X-class flare is also indicated (\textit{green triangle}).}
\label{fig:vzdl2}
\end{figure*}

As this manuscript has detailed in \S\,\ref{sec:decomposing}, the excess current-carrying topology for magnetic helicity and magnetic winding were shown to be promising candidates as flare precursors. However, these metrics alone do not provide information on whether the fluxes associated with these quantities seen at the photospheric level are due to new magnetic field emerging from within the solar interior, or whether it is because of a disturbance in the solar atmosphere that causes existing field to be `pushed' to the photospheric level and/or below. The significance of knowing this information is that it may allow more interpretation/understanding of the signals (significant spikes) produced by the ARTop time-series. To this end, the current-carrying components of the delta measures for winding and the helicity can be combined with the line-of-sight velocity to form, \vzdl, and \vzdh\ that quantify the speed and direction of magnetic topology input at the photosphere. Whilst, the quantities \dLc\ and \dHc\ are themselves fluxes (rates of input), they represent horizontal motion of field lines, so this quantity combines information about the vertical rise of the fluid transporting the field, with this information about the in-plane motion. Large spikes (dips) in these quantities indicate rapidly emerging (submerging) fluxes at the photospheric level.

In Figure \ref{fig:vzdl} we show two time series of the quantity \vzdl to highlight some of its properties. For AR\,11158 (top panel) there are multiple emergence and submergence events that follow each other in quick succession during the build-up phase (i.e. Time $<60$\,hours or with a larger time gap $37$ and $42$ hours). The submergence events are indicative of motions occurring within the solar atmosphere that push topology down to the photospheric level, indicating that some of the topology build-up seen in Figure\,\ref{fig:dl} is could be due magnetic structure already emerged through the photosphere. Similarly, in the \dLc\ time-series for AR\,12673 there are multiple pairs of opposing-sign spikes some coincident with the X-class flares or (temporally) containing M-class flares. Spatial maps for \vzdl\ are shown for an example submergence and re-emergence of the first X-class flare in the AR\.122673 time series (whose spatial location is indicated with a\textit{green plus}). It is clear from these successive maps that material is pushed to the photospheric level or lower in a localised region around $\left(114^{\circ},-8^{\circ} \right)$ by this flare, which then rebounds and pulls more photospheric material/field into the solar atmosphere, in the vicinity of the second X-class flare (location indicated by the \textit{green triangle}). These events are discussed in more detail in \S\,\ref{sec:12673}.

These phases of emergence and submergence causing winding topology to be seen at the photospheric level are likely separate or interconnected structures where the emergence/submergence of one structure leads to buffeting of magnetic field, causing neighbouring structures to emerge and submerge. As is highlighted in the AR\,11158 \vzdl\ time-series, it appears that a period of emergent topology is required to induce flaring, and so focusing upon the net sign of emergence/submergence may provide additional insight to the likelihood of flaring within flare prediction models. 

\subsection{Spatial Significance of Temporal Spikes}\label{sec:spike_timing}
The results from Figures\,\ref{fig:dl} and \ref{fig:vzdl} indicate that excess current-carrying topology quantities (e.g. \dLc) have potential to be precursive metrics for flaring, provided enough complex magnetic topology has been built up in the solar atmosphere. The variation in time-frame for these precursors ranges between virtually immediate to potentially providing several hours of warning about the possibility of eruptive events taking place. This subsection analyses the spatial distribution of these two types of warnings, focusing on some of the X--class flares for the active regions given in Table\,\ref{tab:sharps}.

\subsubsection{AR\,11158: Spatially Meaningful Spikes}
\begin{figure*}
\centerline{\includegraphics[width=\textwidth]{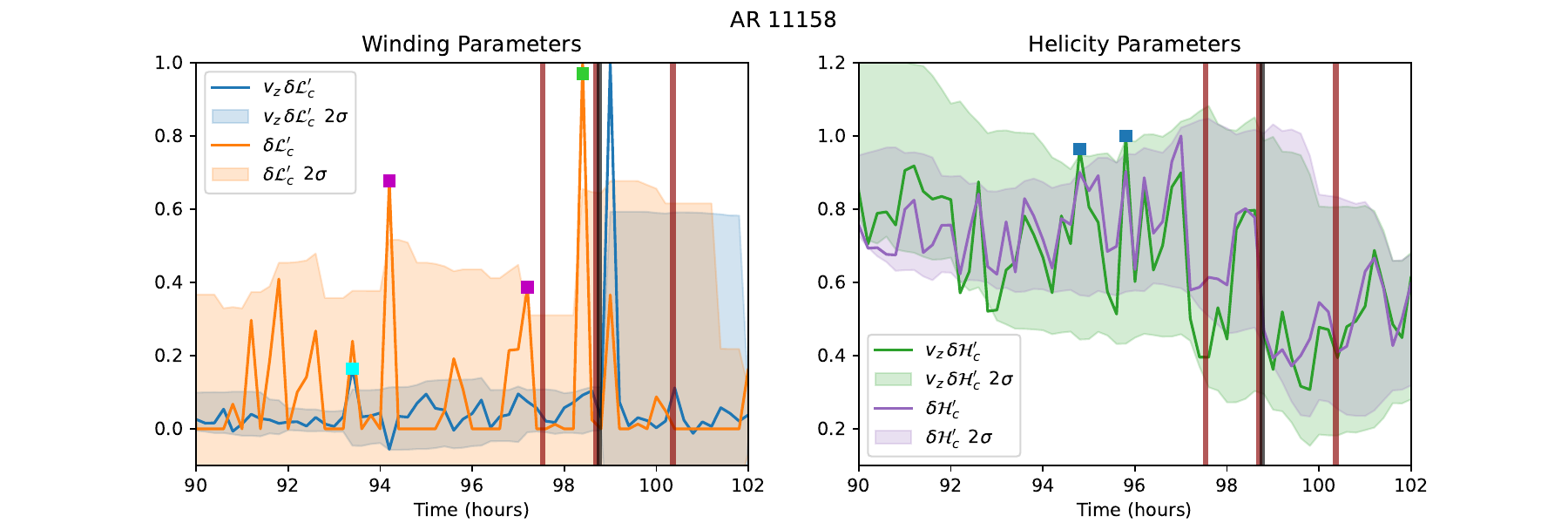}}
\caption{\textit{Left}: Time-series plots for \vzdl\ (\textit{blue}) and \dLc\ (\textit{orange}) from the same period as Figure\,\ref{fig:velcomp} with the corresponding $2\sigma$ envelopes denoted by the shaded regions of the same colours. \textit{Right}: Time-series plots for \vzdh\ (\textit{green}) and \dHc\ (\textit{magenta}) from the same period as \textit{left} with the corresponding $2\sigma$ envelopes denoted by the shaded regions of the same colours. The X2.2 flare is indicated by the \textit{black} vertical line and the three CMEs detected by ALMANAC are shown in \textit{red}.}
\label{fig:x22flare}
\end{figure*}
\begin{figure*}
\centering
\subfigure{\includegraphics[width=0.4\textwidth]{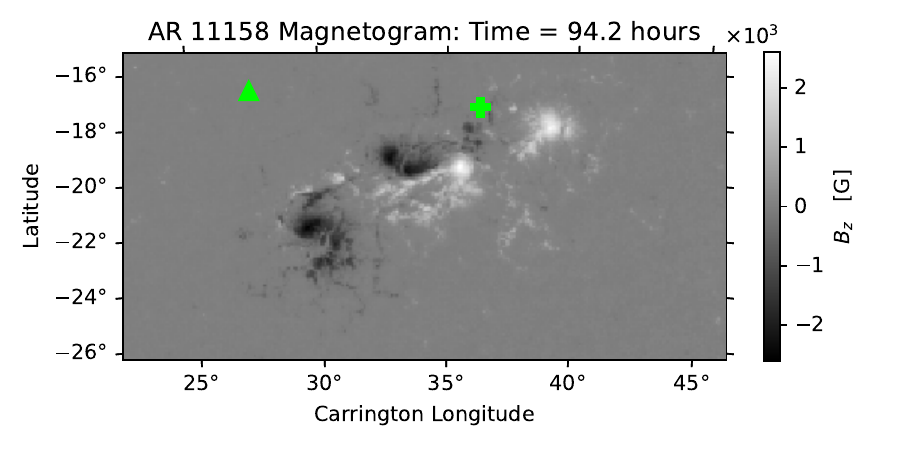}}
\subfigure{\includegraphics[width=0.8\textwidth]{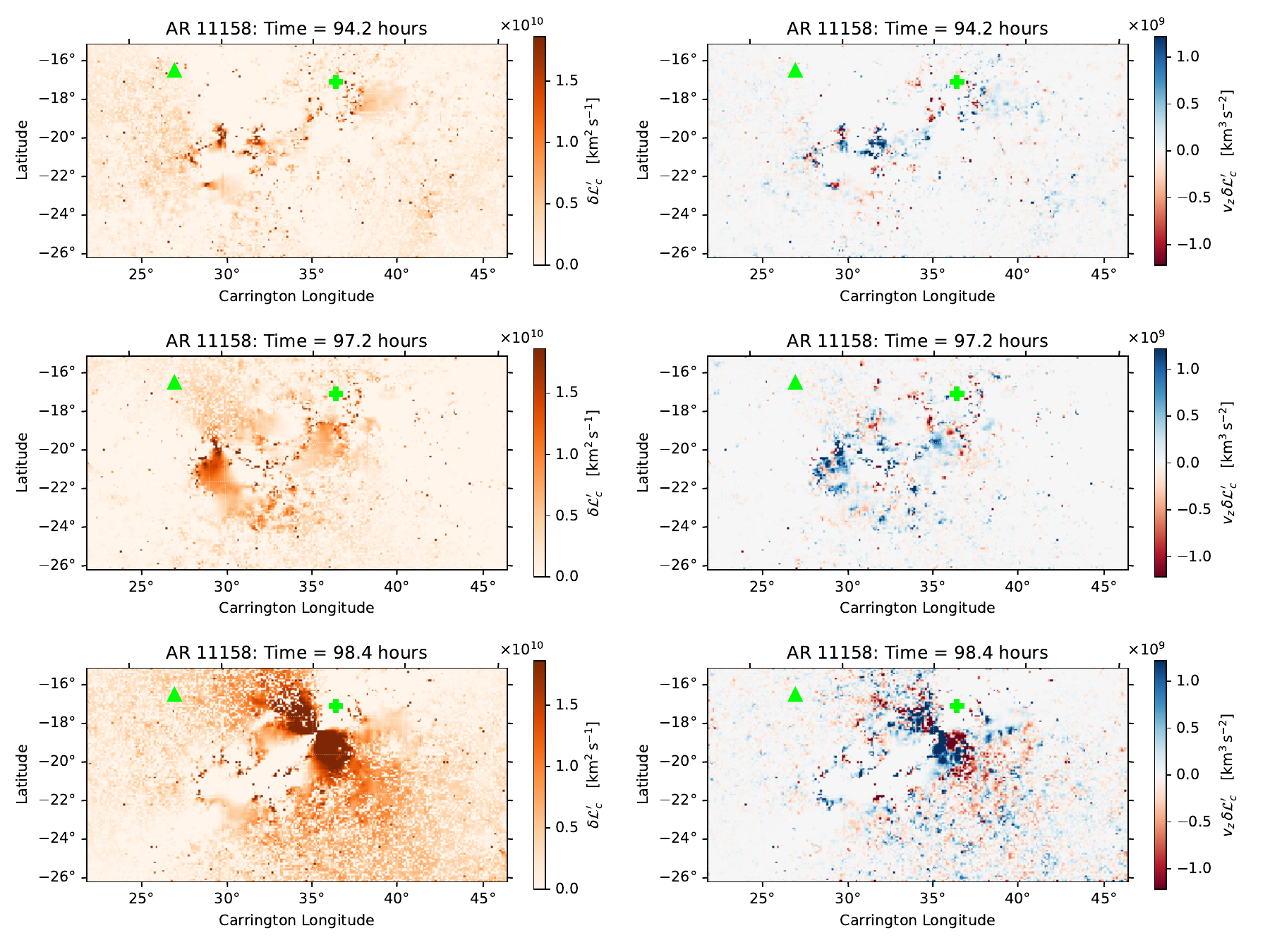}}
\caption{Spatial maps for \dLc\ (\textit{left}) and \vzdl\ (\textit{right}) corresponding to the spikes denoted by the \textit{magenta} and \textit{green squares} in Figure\,\ref{fig:x22flare}. The positions of the first and second CMEs detected by ALMANAC are shown by the \textit{green triangle} and the \textit{green plus}, respectively. The second (third) row corresponds to the surface maps immediately before the first (second) eruption. The magnetogram data from this period is shown for spatial reference.}
\label{fig:377xflaremap}
\end{figure*}

\begin{figure*}
\centering
\subfigure{\includegraphics[width=0.4\textwidth]{SHARP377_X_flare_Bz.pdf}}
\subfigure{\includegraphics[width=0.8\textwidth]{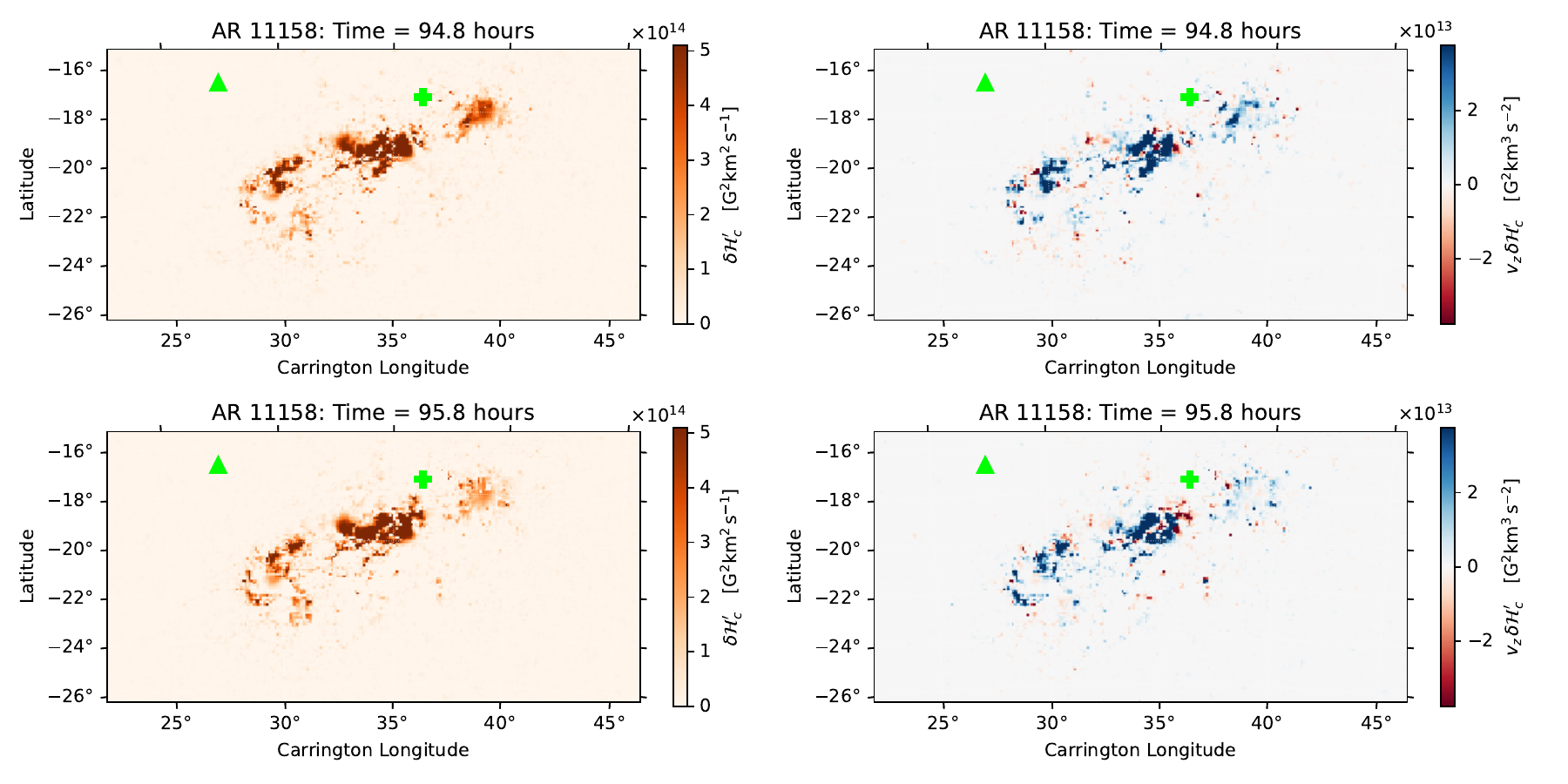}}
\caption{Spatial maps for \dHc\ (\textit{left}) and \vzdh\ (\textit{right}) corresponding to the spikes denoted by the \textit{blue squares} in Figure\,\ref{fig:x22flare}. The top central panel is the $B_z$ magnetogram at $t=$ shown for context. The positions of the first and second CMEs detected by ALMANAC are shown by the \textit{green triangle} and the \textit{green plus}, respectively.}
\label{fig:377xflareHmap}
\end{figure*}

\begin{figure*}
\centerline{\includegraphics[width=0.6\textwidth]{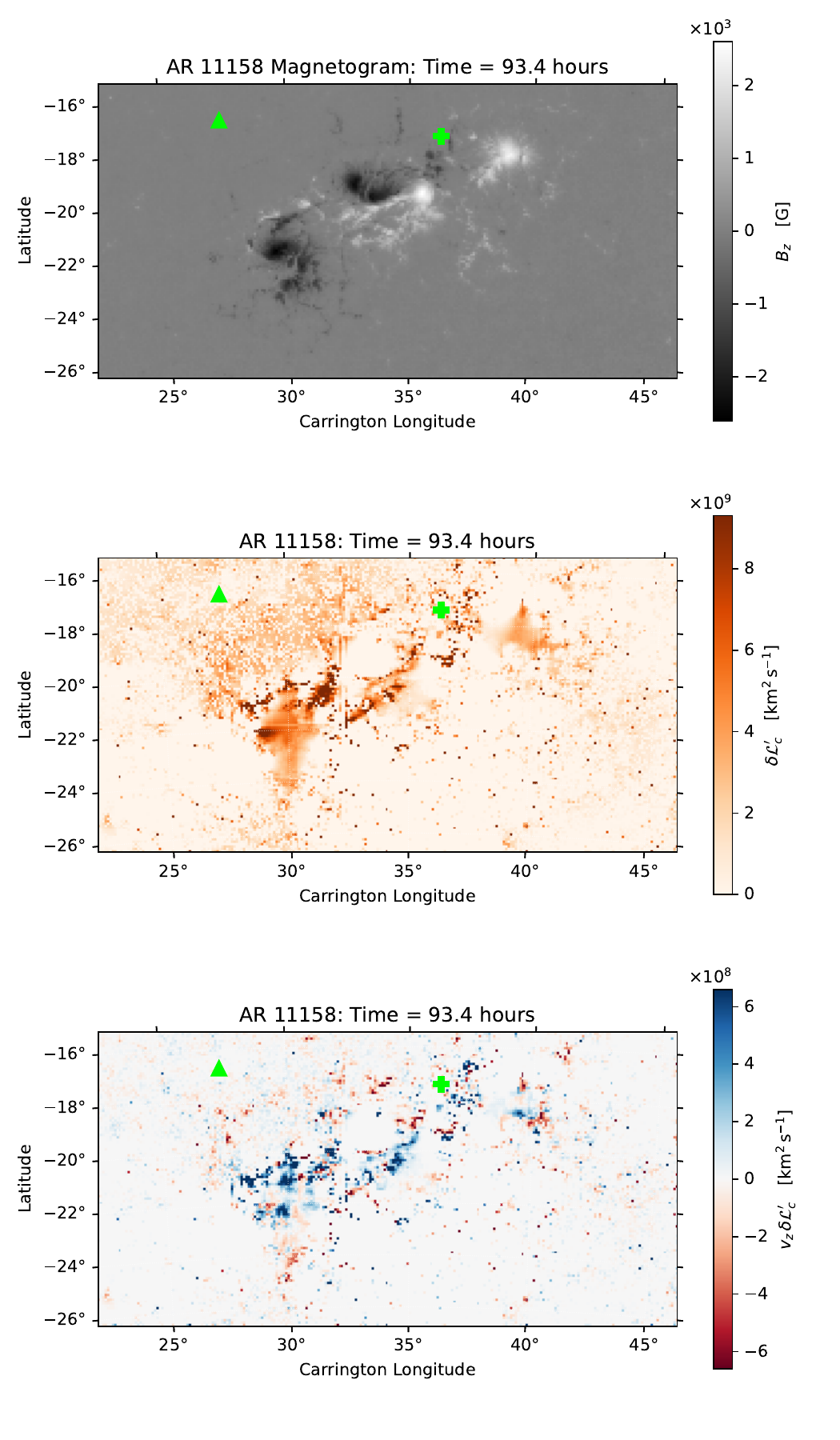}}
\caption{The same as Figure\,\ref{fig:377xflaremap} but for the spikes seen in \dLc\ and \vzdl\ at Time = 93.4\,hours, as is denoted by the \textit{cyan square} in Figure\,\ref{fig:x22flare} with the magnetogram data shown for spatial reference.}
\label{fig:377vzdlspike}
\end{figure*}

In Figure\,\ref{fig:x22flare}, time-series are presented for the current-carrying parameters for both winding and helicity prior to the X2.2 flare and a coincident CME that is detected in LASCO data. Furthermore, within the 12-hour period shown, ALMANAC detected two additional CMEs, the first of which is preceded by significant \dLc\ spikes seen at Time $\approx 94 \text{ and } 97$\,hours (\textit{magenta squares}). The first spike of the two indicated by magenta squares here is caused by a submergence event (negative \vzdl\ at this time), indicating that a change in atmospheric topology caused magnetic field to be pushed down to the photospheric level, highlighting how \vzdl\ may be utilised to infer/capture information about the topological behaviour above the photosphere. There is a spike (green square) at $98.4$ hours which just precedes the coincident flare and CME. There is also an earlier spike in \vzdl\, about one hour before the first \dLc\ spike that indicates an emergence event. For the helicity parameters, \vzdh\ exhibits two spikes between 94 and 96\,hours (\textit{blue squares}), whilst in the hour leading up to the first CME, \dHc\ has a sharp decrease before the first, small CME takes place, with the input rate being somewhat constant before this. A similar trend is also seen with the X2.2 flare and associated CME, though the decline is coincident with the two eruptions in this example.

In Figures\,\ref{fig:377xflaremap} - \ref{fig:377xflareHmap}, the surface maps for three \dLc\, and two \vzdh\ spikes highlighted in Figure\,\ref{fig:x22flare} are shown. The CME at Time $\approx 97.5$\,hours and the event composing an X2.2 flare and CME shortly afterwards are denoted by the \textit{green triangle} and \textit{green plus}, respectively. The two first \dLc\ spikes occur prior to the first small CME captured by ALMANAC (\textit{green triangle}), the spatial maps of the fluxes \dLc\ and \vzdl\ these times are shown in rows 1 and 2 of Figure \ref{fig:377xflaremap}). We see the primary input of topology occur in the vicinity of $\left(29^{\circ},\,-20^{\circ}\right)$ just to the right and below the location of the first CME (triangle). The CME captured by ALMANAC propagates in a northerly direction across the solar disk, which along with the centre of mass for the first frame of detection, is why the authors believe these winding signatures contributed to the first eruption. In figure \ref{fig:377xflareHmap}, the \dHc\ and \vzdh\ input rates at the photosphere (temporally between the first two \dLc\ maps of Figure\,\ref{fig:377xflaremap}) indicate that topology is not only being built-up at $\left(29^{\circ},\,-21^{\circ}\right)$, where the strong winding flux up was seen, but also either side of the location of the later X2.2 flare and CME, with a much stronger contribution in \vzdh\ at $\left(35^{\circ},\,-20^{\circ}\right)$. Chronologically, the last maps (bottom panels of Figure\,\ref{fig:377xflaremap}) reveal large-scale input of topology in the vicinity of the X2.2 flare and CME. The winding contribution for this event has been well documented in \citep{mnraspaper}.


Solely analysing \dLc\ and \dHc\ in this instance indicates some precursive warning on the potential for eruptions with up to $\approx3$\,hours lead-time for the small CME and X2.2 flare events. However, the inclusion of \vzdl\ provides further advanced warning. Prior to Time = 94\, hours, there are four spikes in \dLc -- the last of which is denoted by the \textit{cyan square} in Figure\,\ref{fig:x22flare} -- that do not exceed their $2\sigma$ envelope due to previous signatures (although only just). The last of these (\textit{cyan square}) is associated with strong $v_z$ values, and subsequently \vzdl\ is able to provide an additional 0.8\,hours warning due to the low velocities associated with the other spikes resulting in a smaller $2\sigma$ envelope. We see in Figure \ref{fig:377vzdlspike}, these occur in the same location as the spikes in the first two rows of Figure\,\ref{fig:377xflaremap} and can be attributed to the first CME event (\textit{green triangle}). These events occur up to 7 hours before the CME indicating some earlier warning could be detectable in the time-series. 

\subsubsection{AR\,12673: Destabilising Events}\label{sec:12673}
\begin{figure*}
\centerline{\includegraphics[width=\textwidth]{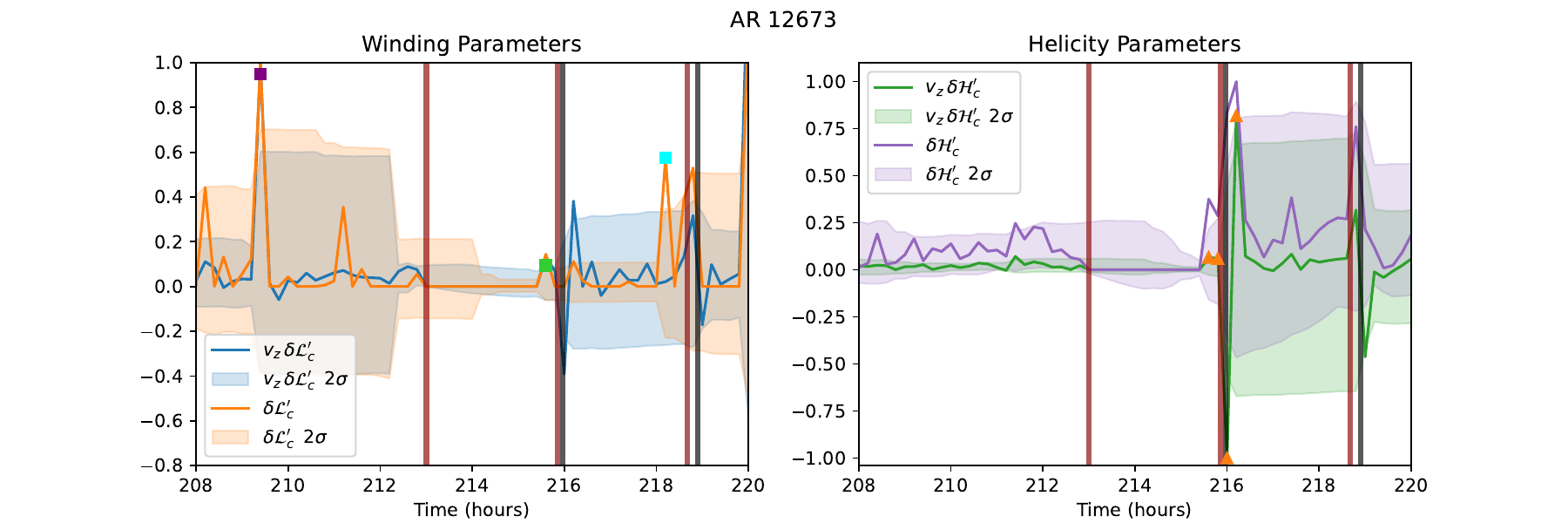}}
\caption{\textit{Left}: Time-series plots for \vzdl\ (\textit{blue}) and \dLc\ (\textit{orange}) for the hours leading to, and including, the X2.2 and X9.3 flares (Times $\approx 216 \text{ and } 219$\,hours) with the corresponding $2\sigma$ envelopes denoted by the shaded regions of the same colours. \textit{Right}: Time-series plots for \vzdh\ (\textit{green}) and \dHc\ (\textit{magenta}) from the same period as Figure\,\ref{fig:velcomp} with the corresponding $2\sigma$ envelopes denoted by the shaded regions of the same colours. The X-class flares are indicated by the \textit{black} vertical lines and the three CMEs detected by ALMANAC are shown in \textit{red}.}
\label{fig:x93flare}
\end{figure*}
\begin{figure*}
\centering
\subfigure{\includegraphics[width=0.4\textwidth]{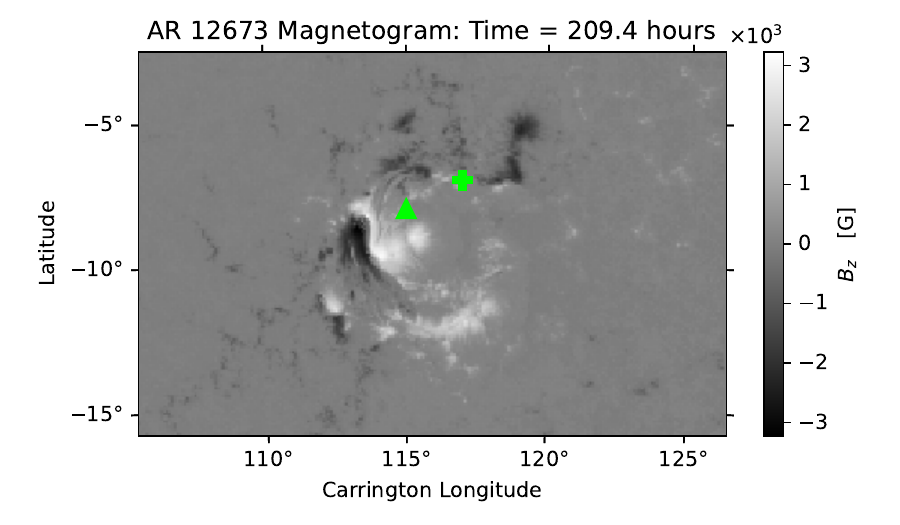}}
\subfigure{\includegraphics[width=0.8\textwidth]{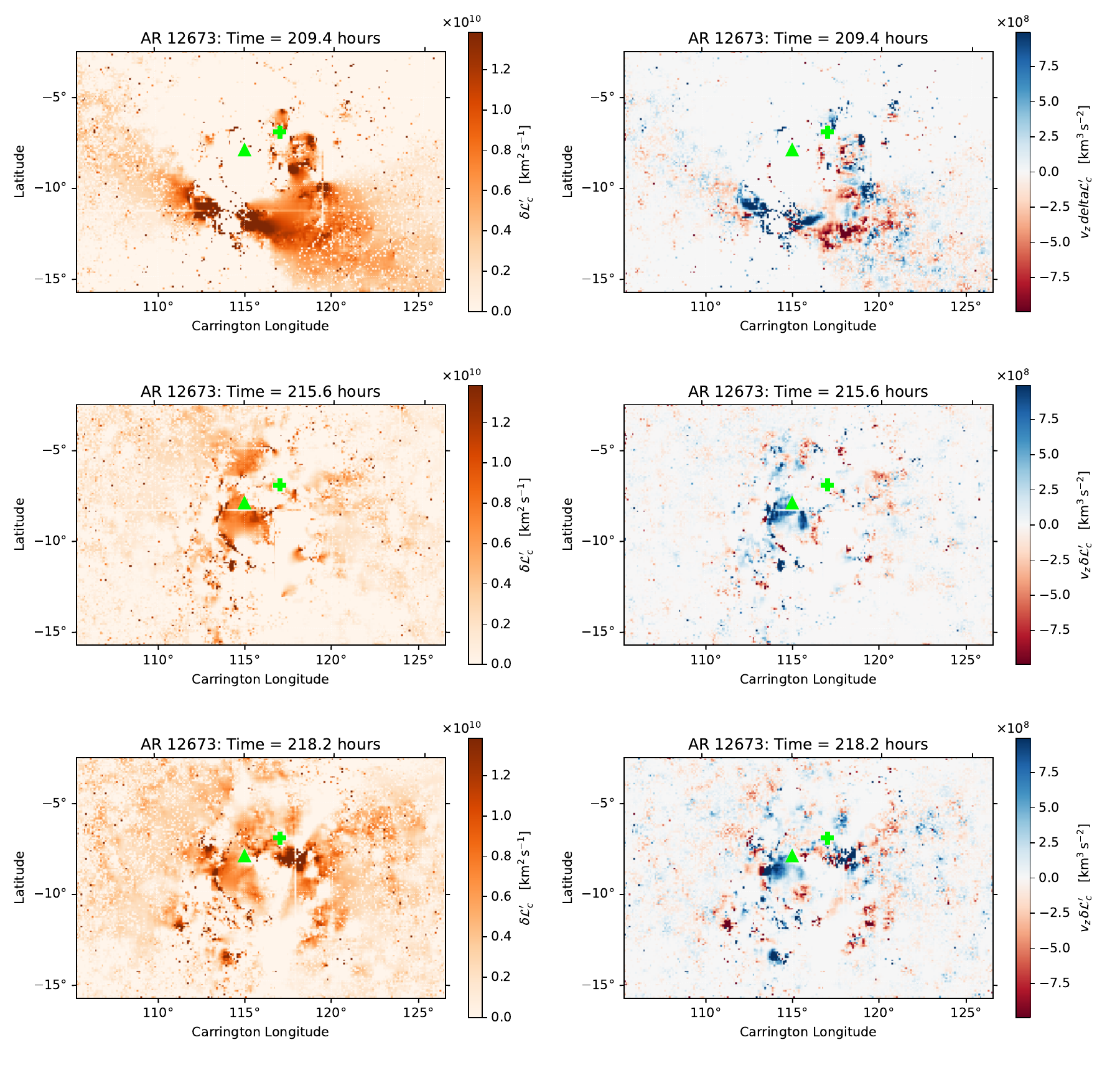}}
\caption{Three snapshots of the \dLc\ and \vzdl\ spatial maps from the times corresponding to the three winding spikes that are highlighted by the \textit{squares} in Figure\,\ref{fig:x93flare}. The position of the X2.2 flare, which erupts just after snapshot 2, is denoted by the \textit{lime green plus}, whilst the X9.3 flare position (erupts after snapshot 3) is shown by the \textit{lime green triangle}. The magnetogram data from this period is shown for spatial reference.}
\label{fig:7115xflaremap}
\end{figure*}

In Figure\,\ref{fig:x93flare}, large coincident spikes in \dLc, and \vzdl\ can be seen (\textit{magenta square}), which precede a CME detected by ALMANAC by $\approx4$\,hours. Second, coincident spikes are seen shortly before the X2.2 flare and CME (\textit{green square}) and a solitary spike in \dLc\ (\textit{cyan square}) before the slowly rising filament eruption that immediately precedes, and potentially triggers, the X9.3 flare are also seen. For the helicity, there is again some warning to the first CME, though much later and smaller in magnitude than the winding signature ($\approx 209.5$\,hours compared to $\approx 211$\,hours). Interestingly, \dHc\ has a spike exceeding its $2\sigma$ envelope that is co-temporal with the winding signature spikes, which then further increases after the X2.2 flare and CME (\textit{orange triangles}). For \vzdh\ (and \vzdl), there is a prominent submergence during the eruption which is immediately followed by an emergence with both these events exceeding their envelopes in \vzdh\ and \vzdl\ (\textit{orange triangles}).

Following the magnetic field model of \citet{price19} for AR\,12673, there are two separate flux ropes that reside close to one another that 1) closely match the submerging/re-emerging winding signatures, and 2) are the structures that are responsible for the X2.2 flare and CME. The first row of Figure \ref{fig:7115xflaremap} shows the winding maps for \dLc\ and \vzdl\, which reveal that the large spike denoted by the \textit{magenta square} in Figure\,\ref{fig:x93flare} is the result of a large region of strong topology primarily emerging at the photospheric level. Following the modelling of this region by \citet{price19}, the strong emergence seen here between $\left(112^{\circ}-117^{\circ},\,-13^{\circ}- -10^{\circ}\right)$ is a low-lying null point that emerges between two pre-existing, overarching flux ropes, further increasing the complexity of the magnetic field for the active region. Thus, when additional topology emerges beneath the flux ropes, as shown in the middle row just prior to the X2.2 flare and CME, and then in the bottom row just prior to the X9.3 flare, the events follow shortly after as the emerging structures destabilise an already complex field. 

\subsubsection{AR\,11302: Consequential and Inconsequential Spikes}
\begin{figure*}
\centerline{\includegraphics[width=\textwidth]{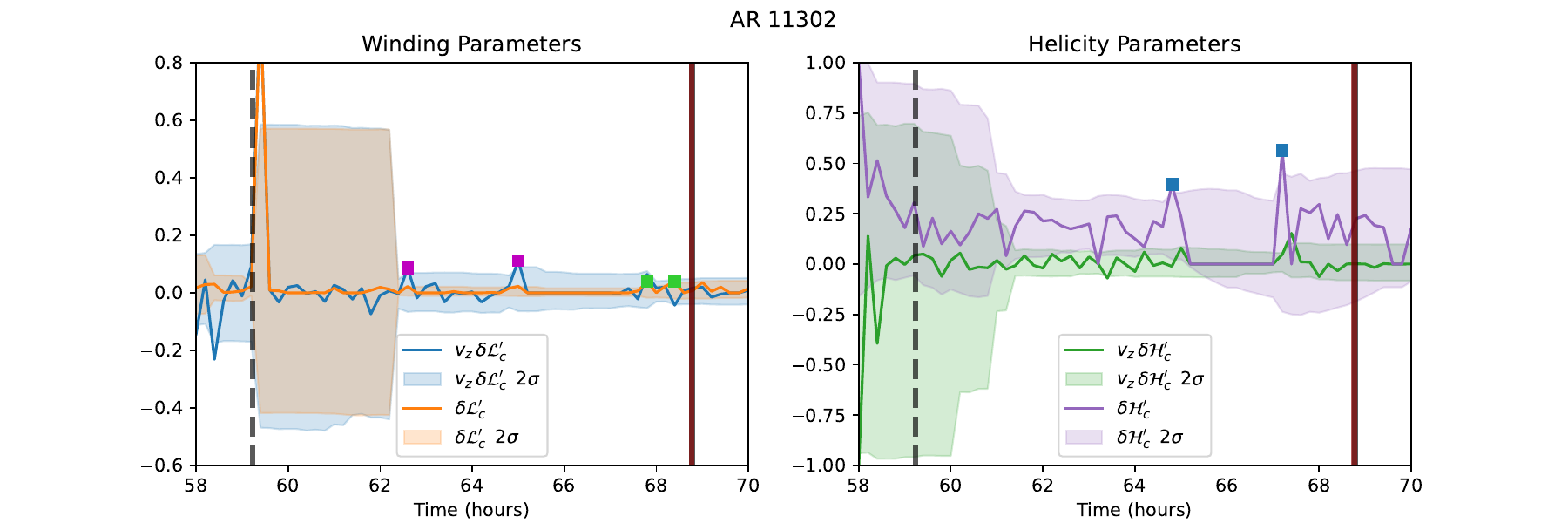}}
\caption{\textit{Left}: Time-series plots for \vzdl\ (\textit{blue}) and \dLc\ (\textit{orange}) in region AR12673 for the hours leading to, and including, the X1.9 flare with the corresponding $2\sigma$ envelopes denoted by the shaded regions of the same colours. \textit{Right}: Time-series plots for \vzdh\ (\textit{green}) and \dHc\ (\textit{magenta}) from the same period as Figure\,\ref{fig:velcomp} with the corresponding $2\sigma$ envelopes denoted by the shaded regions of the same colours. The X1.9 flare is indicated by the \textit{solid black} vertical line, whilst the M1.9 flare is shown as a \textit{dashed black} vertical line, and the solitary CME detected by ALMANAC is shown in \textit{red}.}
\label{fig:x19flare}
\end{figure*}
\begin{figure*}
\centering
\subfigure{\includegraphics[width=0.45\textwidth]{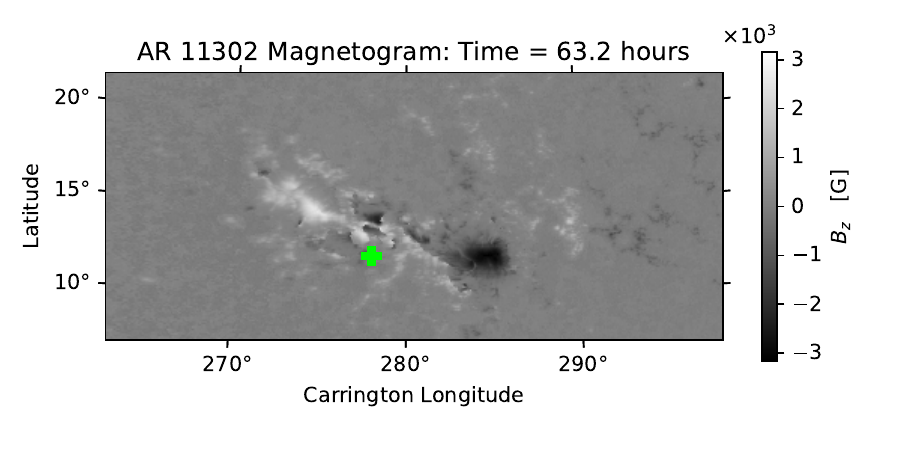}}
\subfigure{\includegraphics[width=0.8\textwidth]{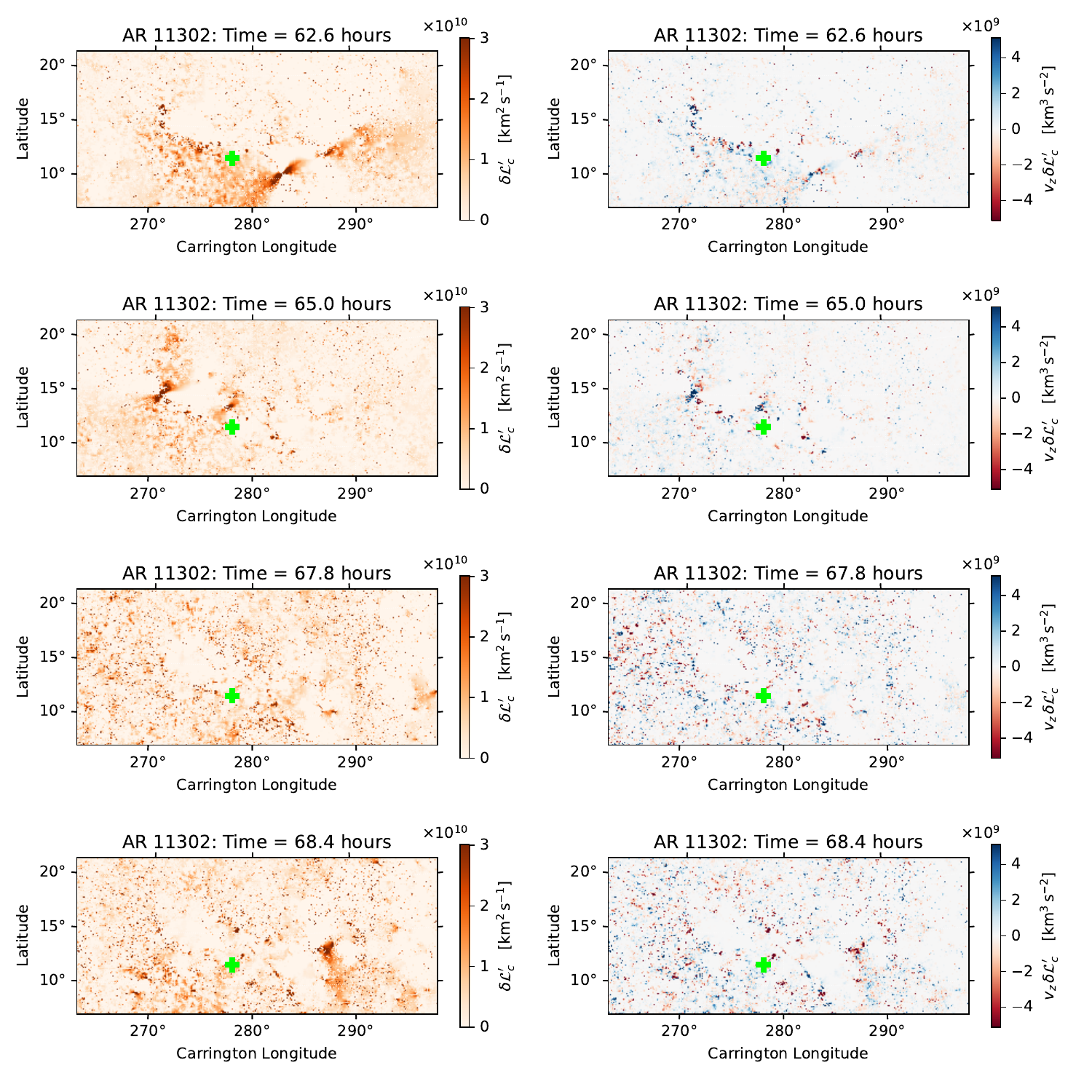}}
\caption{Four sets of snapshots of the \dLc\ and \vzdl\ spatial maps from the times corresponding to the four winding spikes that are highlighted by the \textit{squares} in Figure\,\ref{fig:x19flare}. The position of the X1.9 flare, which erupts just after snapshot 4, is denoted by the \textit{lime green plus}. The magnetogram data from this period is shown for spatial reference.}
\label{fig:892xflaremap}
\end{figure*}
\begin{figure*}
\centering
\subfigure{\includegraphics[width=0.4\textwidth]{SHARP892_X_flare_Bz.pdf}}
\subfigure{\includegraphics[width=0.8\textwidth]{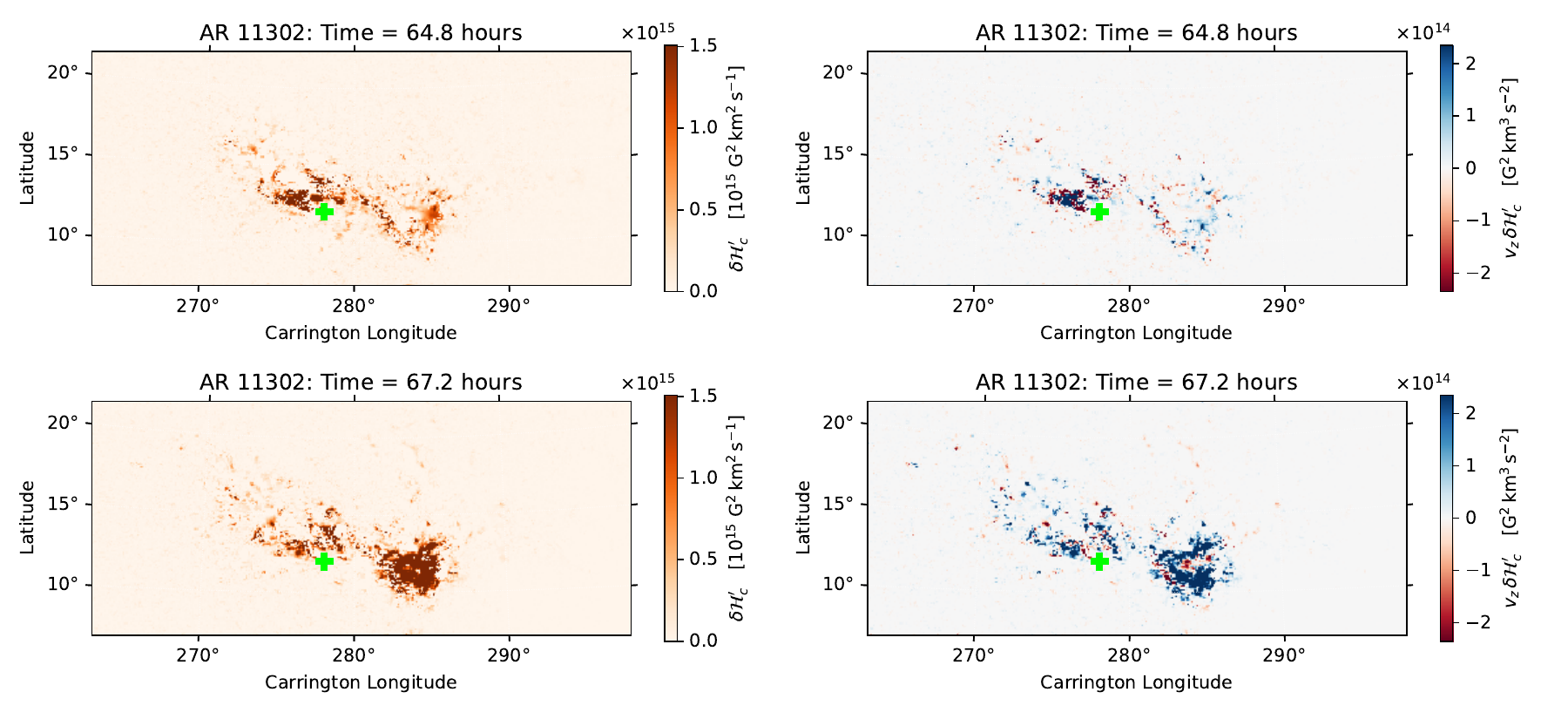}}
\caption{Four successive snapshots of the \dHc\ and \vzdh\ spatial maps temporally surrounding the X1.9 flare and CME (\textit{orange triangles} in Figure\,\ref{fig:x19flare}). The position of the X2.2 flare, which erupts just after snapshot 2, is denoted by the \textit{lime green plus}, whilst the X9.3 flare position (erupts after snapshot 3) is shown by the \textit{lime green triangle}. The magnetogram data from this period is shown for spatial reference.}
\label{fig:892xflareHmap}
\end{figure*}
\begin{figure*}
\centerline{\includegraphics[width=0.7\textwidth]{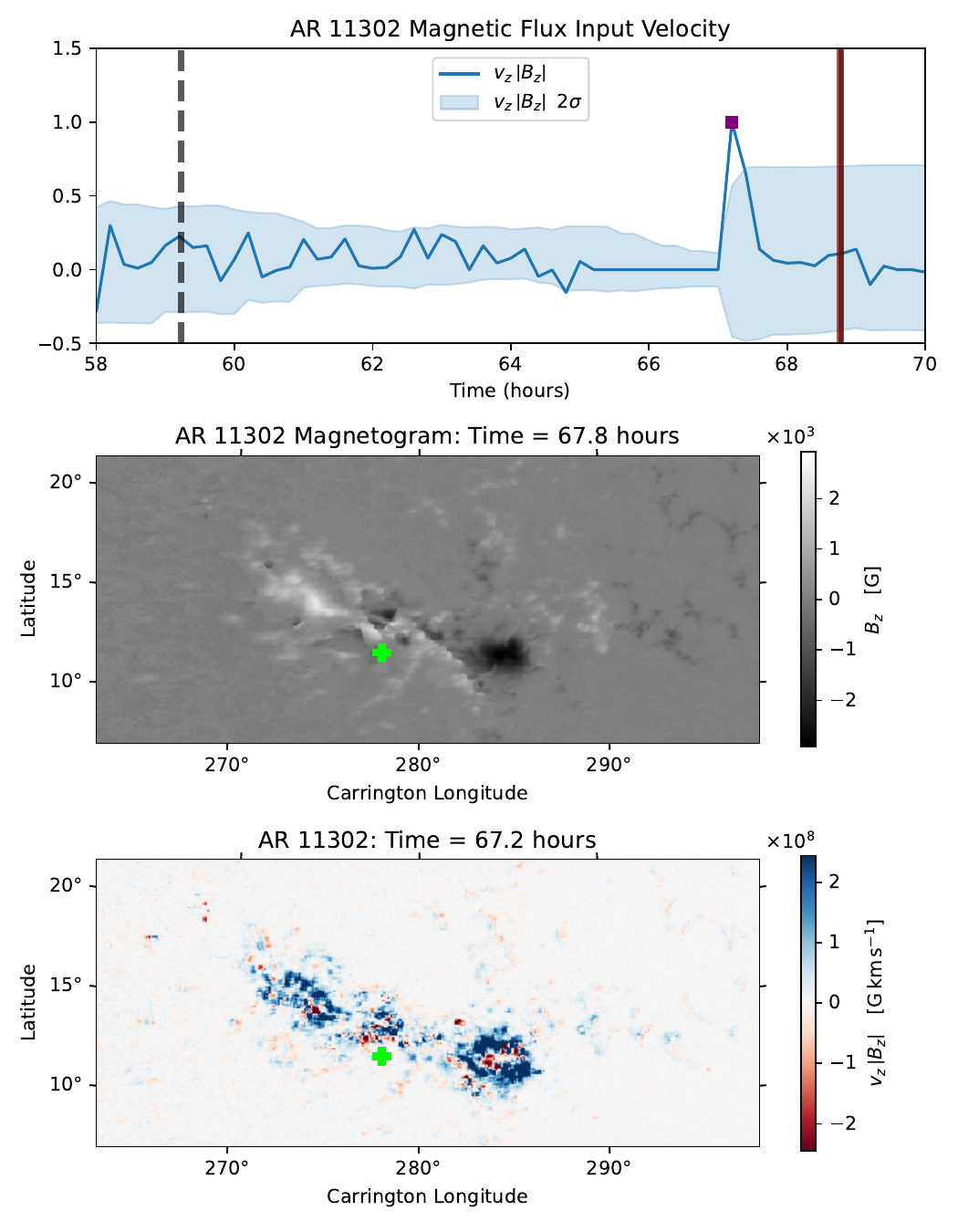}}
\caption{\vzbz\ time-series of AR\,11302 in the hours leading up to the X1.9 flare. The magnetogram and \vzbz\ maps are plotted for the time indicated by the \textit{magenta square}.}
\label{fig:892vzbzmap}
\end{figure*}

For AR\,11302, we focus upon the X1.9 flare (approx $68.5$ hours). In the hours leading up to this event, the topological input has numerous significant potential contributions, as can be seen, for example, between Time $\approx 62$-$68$\,hours in both the \dLc\ and \dHc\ plots (Figure\,\ref{fig:x19flare}). Additionally, unlike the other regions showcased in this manuscript, the X1.9 flare at $\approx69$\,hours is preceded by relatively small spikes in both the winding and the helicity parameters when, compared to the rest of the time-series (see Figure\,\ref{fig:dl}). We note these plots also exhibit relatively flat gradients during this period. The spatial distributions associated with these four spikes, two each in \vzdl\ (\textit{magenta squares} in Figure\,\ref{fig:x19flare}) and \dLc\ (\textit{green squares} in Figure\,\ref{fig:x19flare}) are shown spatially in Figure\,\ref{fig:892xflaremap}. For the first two spikes, shown in the first two rows of Figure\,\ref{fig:892xflaremap}, we see spatially that the concentrations of winding and helicity parameters are within the locality of X1.9 flare for the first two spikes (in the \vzdl\ time-series). By contrast, the spikes in \dLc\, denoted by the \textit{green squares} immediately before the eruption, do not seem to have a spatial correlation with the event, as shown in the bottom two rows of Figure\,\ref{fig:892xflaremap}. Furthermore, it appears they are only seen as significant spikes due to the paucity of current-carrying topological input in the period prior.

Focusing upon the smaller spikes; in the first case (row 3) there are no strong spatial signatures in \dLc\ or \vzdl. By contrast, in the second case (row 4) there is a signature seen at $\left( 287^{\circ},\,13^{\circ} \right)$ (Figure\,\ref{fig:892xflaremap}, snapshot 4) that shows some spatial agreement with the strong helicity inputs show in Figure\,\ref{fig:892xflareHmap}. These correspond to the spikes (\textit{blue squares in Figure\,\ref{fig:x19flare}}) seen in \vzdh\, which do appear more significant in relation to the rest of the time-series. Looking at the extended flare series in Figure\,\ref{fig:dl}, there are C- and M-class flares between 70-80\,hours, and so it is possible that the second patch of helicity centred at approximately $\left(283^{\circ},\,12^{\circ}\right)$ contributed to these later events (as potentially did the winding) and the temporal coincidence with the X1.9 flare is down to chance, a warning for any potential predictive method based off these quantities.

 Then, if we consider the combination of parameters \vzbz\ (Figure\,\ref{fig:892vzbzmap}), a measure of speed of input/removal of flux, we see a spike at 67.2\,hours (\textit{magenta square}) a significant spike in the rate. As with the winding signatures seen at Time = $33$\,hours (Figure\,\ref{fig:vzdl}), the input of flux is centred around the location of the X1.9 flare and so the spikes seen at both these times are likely meaningful due to the spatiotemporal proximity to eruptions. Both these observations indicate that in carefully monitoring related quantities, context may be added to the spikes seen in the winding series.

The main implications of this subsection is that the winding, and to a lesser degree helicity spikes in the time-series are likely a good indicator for the onset of solar eruptions within a given region of the Sun. When analysed spatially, the signatures of these spikes show strong spatial agreement with the eruptions, as has also been demonstrated for CMEs by \citet{mnraspaper}. However, the magnitude of the spikes alone, as highlighted by the results for AR\,11302, are not likely to be indicative of the magnitude or number of (sympathetic) eruptions.

\subsection{Time-Series Kurtosis: Gauging the Importance of Spikes}\label{sec:kurtosis}
\begin{figure*}
\centerline{\includegraphics[width=0.7\textwidth]{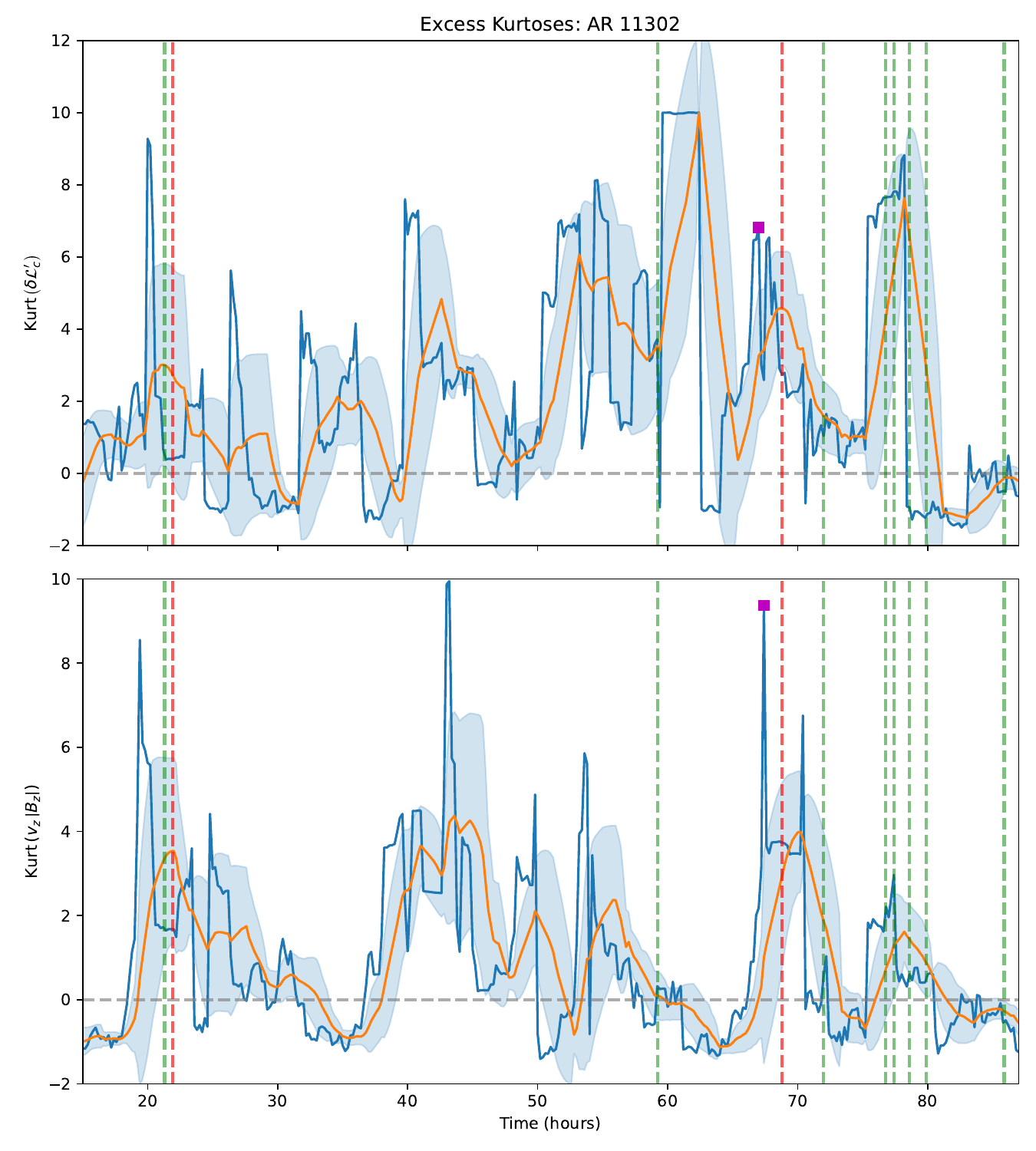}}
\caption{Excess kurtosis plots for \dLc\ and \vzbz\ are shown in \textit{blue}, with their respective 3-hour running means (\textit{orange}) and $1\sigma$ envelopes \textit{shaded blue} indicated. The $y=0$ line (\textit{dashed gray}) indicates the value for a normal distribution. As with previous figures, M- and X-class flares are indicated by the dashed vertical \textit{green} and \textit{red} lines, respectively.}
\label{fig:kurtosis}
\end{figure*}

As this manuscript highlighted in Figure\,\ref{fig:dl}, AR\,11302 exhibits spikes that precede X-class flares which, when compared to the $2\sigma$ envelope, are significant, yet when compared to other spikes in their respective time-series, the magnitudes of the spikes appear relatively inconsequential, especially when compared to the X-class magnitude flare that follows shortly after them. In this subsection we further explores these events using a different metric to see if there could be some indication of their likely magnitude.

Having explored the spatial distributions of these small spikes in \S\ref{sec:spike_timing} (e.g. Figures\,\ref{fig:892xflaremap} and \ref{fig:892xflareHmap}), it is difficult to determine from visual inspection whether an input of new topology could contribute to a magnetic reconnection event in the form of a large flare, by comparison to the much more pronounced examples seen for AR\,11158 and AR\,12673. Whilst this may be something a neural network could be trained to determine -- something that is beyond the scope of this study -- we feel further statistics are likely required that assess the recent behaviour of an active region. One we explore here is the excess kurtosis of a given time-series, calculated utilising a kernel width of 3-hours. Examples of this quantity calculated for the time-series of \dLc\ and \vzbz\ for AR\,11302 and are shown in Figure\,\ref{fig:kurtosis} during the same observation window as is shown in Figure\,\ref{fig:dl}. Values greater than $0$ indicate that there is a larger skew in the recent behaviour of the given series towards the tails of the series values, by comparison to a normal distribution. That is to say the series is exhibiting relatively extremal behaviour, something that could potentially indicate a system destabilizing and having an increased propensity that some form of eruption may take place.

In these two instances of the excess kurtosis time-series shown in Figure\,\ref{fig:kurtosis}, its value can be seen to periodically switch between being leptokurtic (excess $> 0$) and platykurtic (excess $< 0$), with both being leptokurtic dominant suggesting that generally, the 3-hour kernel for these quantities have larger tails than a normal distribution. For the X-class flare in AR\,11302, we can see that the relative magnitude of the \dLc\ excess kurtosis spikes prior to this event (\textit{magenta square}) has a comparable magnitude to other spikes seen throughout the time-series; something that is in stark contrast to the relative size of the \dLc\ spikes at the same time (\textit{green triangles} in Figure\,\ref{fig:dl}). It is also worth mentioning that the spike shown here is not the only one that exceeds the $1\sigma$ envelope, as there is a cluster of spikes that exceed this envelope in the period prior to the flare, and so we have only highlighted the largest spike. In the excess-kurtosis for \vzbz\, a similar result to that shown in Figure\,\ref{fig:892vzbzmap} is seen whereby the spike preceding the flare is one of largest seen across the entire observation window, providing further indication that a potentially large eruption may take place.

In summary, evaluating the excess kurtosis of time-series may provide valuable additional weighting to spikes seen in the original time-series when predicting the likelihood of flaring. Individual spikes seen in quantities such as \dLc, provide a glimpse into the instantaneous behaviour of the region, whilst quantities such as the accumulative helicity provide the long-term topology input into a region. The running excess kurtosis complements these statistics by assessing the recent behaviour of the region, and quantifying the ``tailedness'' or extremal nature of its recent behaviour. As we have shown in this subsection, a sudden rise in the excess kurtosis could be a key factor in classifying the likely magnitude of a flare.

\section{Preliminary Investigations into Forecasting Efficacy}\label{sec:results2}
In this section, the focus is on how consistently the quantities outlined in \S\,\ref{sec:results} may be used to predict the likelihood of flaring. These results should be seen only as as a preliminary indication into how frequently spikes in the aforementioned time-series correlate with flare activity, rather than the development of a live predictive diagnostic which is beyond the scope of this study. The aim is to give users an indication to the consistency of each time-series with regards to spikes correlating with flares and also how other parameters like the time-integrated $\delta H$ might be used to classify spikes as meaningful, as discussed in \S\,\ref{sec:results}. 

For this, we first analyse the three example regions; AR\,11158, AR\,11302, and AR\,12673 before utilising a more comprehensive dataset (see Appendix). We define a score, $S$:
\begin{equation} \label{eq:bestPerformer}
    S = \frac{1}{2}\left(X_s + X_f \right),
\end{equation}
where $X_s$ and $X_f$ are the percentage of spikes preceding flares within a number of hours ($X_s$), and the percentage of flares following a spike within a number of hours ($X_f$), respectively. The first quantity, $X_s$, quantifies the likelihood a flare being identified with a given spike (in some given metric) within a given period: is this metric good at indicating flares? The second quantity, $X_f$ attempts to quantify how often a spike can be expected to precede a given flare in a specific period. It is necessary to consider both as some metrics, for example, may generate a vast number of spikes and capture almost all flares, but in doing so too often yield spikes not associated with flares that would render the metric ineffective. Ideally, the measure will be maximised by metrics that only produce meaningful spikes and do so as consistently as possible.

A weighted score is also defined as $S_w$, which accounts for the number of spikes and flares within a region. This metric assigns greater importance to regions where flaring is more prominent, but has no dependency on the magnitude of the events. The weighted score is defined by:
\begin{equation}\label{eq:weightedBestPerfromer}
    S_w = \frac{1}{2}\, \left(\frac{\sum_{i} X_{s_i} N_{s_i}}{\sum_{i} N_{s_i}} + 
                              \frac{\sum_{i} X_{f_i} N_{f_i}}{\sum_{i} N_{f_i}}
                        \right) ,
\end{equation}
where subscript $i$ indicates the SHARP patch. $N_{si}$ and $N_{fi}$ are the total number of spikes and flares for that region, respectively. If $S$ and $S_w$ provide similar results for a quantity across a number of active regions, it is indicative that the quantity is not biased towards predicting flaring for particular regions over others. However, should $S_w$ drastically exceed $S$ for example, this would indicate a metric is outperforming for active regions with more flare activity and we should trust it less for less active regions. 

\subsection{Example Regions}
As outlined in \S\,\ref{sec:results}, the quantities determined from ARTop, namely \dLc, \dHc, \vzdl, \vzdh, and \vzbz\ often have spikes which precede flaring, and so the aim here is to quantify the reliability of spikes in these quantities before an eruption. As with the previous sections, we utilise 3\,hour running means and 2\,$\sigma$ envelopes for all the quantities analysed. As has already been noted in \S\,\ref{sec:results}, some of the earlier spikes in the time-series do not lead to flaring, and this typically seems to be when insufficient helicity accumulation has taken place through the photosphere. Consequently, we also quantify the percentage of spikes preceding flares ($X_s$) and flares preceded by spikes ($X_f$) with and without a cutoff for the helicity accumulation in Table\,\ref{tab:stats1}. When employing the helicity cutoff, if a spike exceeding the $2\sigma$ envelope is seen in, for example \dLc, but the total helicity accumulation for that active region is below the cutoff, then it is considered to be inconsequential and is omitted from the statistics. Finally, we consider $3$, $6$ and $12$ hour search windows over which we correlate spikes and flares; one would expect that with an increasing search window size that $S$ and $S_w$ would also increase, but in that case it is less likely one could expect a direct causal relationship between the spike and the flare. We have seen above that it is possible to correlate spikes spatially within periods of up to $9$ hours, and so the $12$ hour search window here is used to provide some insight into how the metic changes with increased window size.

\begin{table}[htb]
\centering
\caption{Percentage of spikes preceding flares, $X_s$ and flares with spikes preceding them, $X_f$ across the three example regions.}

\begin{tabular}{c}
No helicity cutoff applied.\\
\end{tabular}

\begin{tabular}{c c c c c c c c c}\label{tab:stats1}
\\[-1em] 
\hline
\multirow{2}{*}{Quantity} & \multicolumn{3}{c}{$X_s$} & \multirow{2}{*}{$N_s$} & 
\multicolumn{3}{c}{$X_f$} & \multirow{2}{*}{$N_f$} \\
& 3\,hr & 6\,hr & 12\,hr & & 3\,hr & 6\,hr & 12\,hr & \\
\hline
\dLc  & 0.470 & 0.566 & 0.697 & 236 & 0.556 & 0.756 & 0.848 & 190 \\
\dHc  & 0.380 & 0.502 & 0.653 & 166 & 0.371 & 0.549 & 0.737 & 190 \\
\vzdl & 0.383 & 0.494 & 0.652 & 163 & 0.383 & 0.608 & 0.790 & 190 \\
\vzdh & 0.356 & 0.518 & 0.648 & 156 & 0.349 & 0.574 & 0.749 & 190 \\
\vzbz & 0.402 & 0.512 & 0.608 & 113 & 0.291 & 0.537 & 0.735 & 190 \\

$\mathrm{Kurt}\left( \delta \mathcal{L}_c^{\prime} \right)$  
& 0.428 & 0.551 & 0.644 & 291 & 0.474 & 0.702 & 0.803 & 190 \\
$\mathrm{Kurt}\left( \delta \mathcal{H}_c^{\prime} \right)$  
& 0.359 & 0.512 & 0.646 & 261 & 0.423 & 0.688 & 0.854 & 190 \\
$\mathrm{Kurt}\left( v_z\,\delta \mathcal{L}_c^{\prime} \right)$  
& 0.470 & 0.600 & 0.669 & 265 & 0.401 & 0.721 & 0.842 & 190 \\
$\mathrm{Kurt}\left( v_z\,\delta \mathcal{H}_c^{\prime} \right)$  
& 0.399 & 0.539 & 0.616 & 289 & 0.477 & 0.681 & 0.850 & 190 \\
$\mathrm{Kurt}\left( v_z\,|B_z| \right)$  
& 0.452 & 0.546 & 0.668 & 269 & 0.433 & 0.677 & 0.835 & 190 \\

\hline
\end{tabular}

\begin{tabular}{c}
\\[-0.5em]
Helicity cutoff of $1\times10^{19}$\,Mx$^2$ applied. \\
\end{tabular}

\begin{tabular}{c c c c c c c c c}
\hline
\multirow{2}{*}{Quantity} & \multicolumn{3}{c}{$X_s$} & \multirow{2}{*}{$N_s$} & 
\multicolumn{3}{c}{$X_f$} & \multirow{2}{*}{$N_f$} \\
& 3\,hr & 6\,hr & 12\,hr & & 3\,hr & 6\,hr & 12\,hr & \\
\hline
\dLc  & 0.554 & 0.654 & 0.785 & 207 & 0.545 & 0.728 & 0.814 & 190 \\
\dHc  & 0.469 & 0.631 & 0.769 & 134 & 0.355 & 0.521 & 0.698 & 190 \\
\vzdl & 0.462 & 0.597 & 0.743 & 138 & 0.366 & 0.575 & 0.751 & 190 \\
\vzdh & 0.465 & 0.694 & 0.806 & 118 & 0.327 & 0.540 & 0.716 & 190 \\
\vzbz & 0.564 & 0.688 & 0.791 & 92 & 0.280 & 0.514 & 0.707 & 190 \\

$\mathrm{Kurt}\left( \delta \mathcal{L}_c^{\prime} \right)$  
& 0.519 & 0.658 & 0.738 & 257 & 0.474 & 0.691 & 0.787 & 190 \\
$\mathrm{Kurt}\left( \delta \mathcal{H}_c^{\prime} \right)$  
& 0.400 & 0.569 & 0.714 & 231 & 0.406 & 0.655 & 0.821 & 190 \\
$\mathrm{Kurt}\left( v_z\,\delta \mathcal{L}_c^{\prime} \right)$  
& 0.527 & 0.671 & 0.736 & 239 & 0.379 & 0.688 & 0.803 & 190 \\
$\mathrm{Kurt}\left( v_z\,\delta \mathcal{H}_c^{\prime} \right)$  
& 0.487 & 0.633 & 0.723 & 250 & 0.471 & 0.658 & 0.822 & 190 \\
$\mathrm{Kurt}\left( v_z\,|B_z| \right)$  
& 0.512 & 0.611 & 0.728 & 243 & 0.422 & 0.649 & 0.802 & 190 \\

\hline

\end{tabular}
\end{table}

First we focus on the values of $X_s$ and $X_f$ which are presented in Table\,\ref{tab:stats1} for five quantities and their respective excess kurtoses for the three example regions. We focus on these quantities rather than $S$ and $S_w$ to get an idea of how performance in the two quantities varies. As one would expect, the greater the search duration for a flare (spike) following (preceding) a spike (flare), the larger the percentage of matches seen for $X_s$ ($X_f$). Without a helicity cutoff, the $X_s$ values show that $\approx 35-47$\,\% of spikes are typically followed by a flare within 3\,hours, whilst $\approx 29-56$\,\% of flares are preceded by spikes ($X_f$). As that window is broadened to 6 and then to 12\,hours, $X_s$ and $X_f$ increase with the best scores becoming 69.7\,\% and 85.4\,\%, respectively. However, it is worth noting that only \dLc, and the excess kurtoses have a sufficient number of spikes to be able to `capture' all of the flares from the example regions. When applying a helicity cutoff of $1\times10^{19}\,\mathrm{Mx}^2$, $X_s$ may improve by as much as 19\,\% (\vzbz; 12\,hours), however a more typical improvement is $\approx 7-8$\,\%: spikes are more efficient at correlating with flares. This improvement however, is to the detriment of $X_f$, which typically decreases $\approx 1-2$\,\%: less flares are caught. Additionally, the total number of spikes ($N_s$) decreases, which is a positive outcome for \dLc\ and the excess kurtoses as they still exceed the total number of flares ($N_f$) for the three example regions. This does however further reduce the forecasting efficacy of the other quantities as their $N_s$ values further decrease below $N_f$.

We highlight that there are some quantities for which $N_s<N_f$, that is, fewer predictive spikes than there are flares (relatively high $X_s$ and low $X_f$). These will tend to perform worse with a cutoff, whilst those for which $N_s>N_f$ are candidates for improvement. Intriguingly, whilst is only \dLc\ of the raw quantities that has $N_s>N_f$ marking it as a key quantity in this dataset, the excess kurtosis of all of the explored quantities has this property.

\subsection{Flare Prediction On Large Dataset}
We now turn our attention to the full dataset given in the Appendix. For this, a number of combinations of the parameters which define spikes are tested, namely the running mean period, the standard deviation envelope size outside of which the signal must rise to yield a spike, and the helicity cutoff, which further refines the definition of a meaningful spike or signal (Table\,\ref{tab:params}). The results for the best 10 performing combinations (largest value of the mean $\frac{S + S_w}{2}$) are shown in Tables\,\ref{tab:best3} - \ref{tab:best12}, calculated on  the \FAR\ SHARP regions that undergo flaring. As above, search windows within 3, 6, and 12\,hours of a spike are considered for linking spikes to flares. $N_s$ is also provided for each quantity and parameter combination to show the number of false-positive detections in the \NF\ SHARP regions that have no associated flare activity (we focus on this data shortly).

Immediately, it is apparent that regardless of the period after a spike that is being analysed (i.e. is there a flare?), it is \dLc\ or quantities derived from this that consistently have the best scores for both measures $S$ and $S_w$. As with Table\,\ref{tab:stats1}, $S$, $S_w$, and $\frac{S + S_w}{2}$ all improve in their accuracy as indicators for flaring when the search window is increased to 12\,hours, though the increase here is $\approx20\%$, whereas the three example regions, with numerous, high-energy eruptions a value of $\approx30\%$ was obtained. We also note that scores are slightly down on the values found for our three example regions (\textit{e.g.} $S$ for 3 hours goes from $47\%$ down to $38\%$. We also reflect on the fact the $S_w$ scores are consistently higher than their unweighted counterparts. Many of the regions included have only smaller C-class flares and/or only one or two flares. Subsequently, it is clear the metrics investigated here are providing more meaningful value in more flare rich regions.

On shorter timescales, particularly within 3 hours of a spike (Table\,\ref{tab:best3}), \dLc\ occupies four of the top ten scores, and as with the example shown in Figure\,\ref{fig:kurtosis}, the excess kurtosis calculations also provide comparatively good scores along with \vzdl. As the search window size increases, the parameters in the top ten lists become more varied, with the velocity combinations and their excess kurtoses becoming more prevalent.

\subsubsection{False positives}
Perhaps unsurprisingly due to the nature of $S$ and $S_w$, the majority of top ten scores do not include a helicity cutoff and prefer smaller running mean envelopes, $\sigma$. Increasing these two parameters drastically reduces the number of false-positive detections seen in the non-flaring regions ($N_s$ in Tables\,\ref{tab:best3}-\ref{tab:best12}). For example, \dLc\ calculated (Table\,\ref{tab:best3}) with a running mean of 3\,hours, $\sigma = 1.5$, and a helicity cutoff of $1\times10^{18}\,\mathrm{Mx}^2$ yields scores of $S = 0.311$, $S_w = 0.406$, and $\frac{S + S_w}{2} = 0.359$, whilst only having $N_s = 2672$ across the \NF\ non-flaring region compared to $N_s = 4752$ without the cutoff. However, as outlined earlier, when $S_w$ significantly exceeds $S$ for a metric, it suggests that the metric is outperforming for larger, more active SHARP regions, and should be considered less effective for less active SHARP regions.

We now focus on non-flaring regions whose metric performance is represented by the quantity $N_s$, in effect the number of errors made. It is found that higher helicity cutoffs reduced the number $N_s$ at the cost of significantly lower $S$ and $S_w$ on the flaring regions. The quantities shown here are grossly unreliable for non-flaring regions with thousands of false positives across the dataset. Whilst this manuscript demonstrates that there is clear merit to the topological quantities in regards to spikes preceding flare activity. It may be that a balance could be realised by combining various metrics with more stringent thresholds/cutoffs and employing some form of neural network to predict flaring based on all of the metrics produced, a significant endeavour beyond the scope of this study.

\begin{table}
    \centering
    \caption{List of parameter scans performed.}
    \begin{tabular}{ccc}\label{tab:params}
         Run. Mean. (hr) & $\sigma$ & Helicity Cutoff (Mx$^2$) \\
         \hline
         3  & 1.5 & 0 \\
         6  & 2.0 & $1\times 10^{18}$ \\
         9  & 2.5 & $1\times 10^{19}$ \\
         12 & 3.0 & $1\times 10^{20}$ \\
    \end{tabular}
\end{table}

\begin{table}
    \centering
    \caption{10 best performing statistics in both $S$ and $S_w$ across all \FAR\ SHARP regions analysed with flare activity within 3 hours of a spike.}
    \begin{tabular}{cccccccc}
        \hline
        \multirow{2}{*}{Quantity} & Run. Mean & \multirow{2}{*}{$\sigma$} & Helicity Cutoff & \multirow{2}{*}{$S$} & \multirow{2}{*}{$S_w$} & \multirow{2}{*}{$\frac{1}{2} \left(S + S_W \right)$} & $N_s$ \\
        & (hr) & & (Mx$^2$) & & & & (non-flaring) \\
        \hline
        \dLc & 3 & 1.5 & 0 & 0.376 & 0.408 & 0.392 & 4752  \\
        \dLc & 3 & 1.5 & 1 $\times 10^{18}$ & 0.311 & 0.406 & 0.359 & 2672  \\
        \dLc & 6 & 1.5 & 0 & 0.33 & 0.354 & 0.342 & 4499  \\
        \vzdl & 3 & 1.5 & 0 & 0.325 & 0.344 & 0.335 & 3712  \\
        \kurtvzdh & 3 & 1.5 & 0 & 0.308 & 0.359 & 0.333 & 7942  \\
        \kurtvzbz & 3 & 1.5 & 0 & 0.301 & 0.362 & 0.331 & 7741  \\
        \vzdh & 3 & 1.5 & 0 & 0.321 & 0.337 & 0.329 & 4192  \\
        \dHc & 3 & 1.5 & 0 & 0.303 & 0.348 & 0.326 & 4797  \\
        \dLc & 3 & 2 & 0 & 0.314 & 0.337 & 0.325 & 2655  \\
        \kurtvzdl & 3 & 1.5 & 0 & 0.31 & 0.34 & 0.325 & 7796  \\
        \hline
    \end{tabular}
    \label{tab:best3}
\end{table}
\begin{table}
    \centering
    \caption{10 best performing statistics in both $S$ and $S_w$ across all \FAR\ SHARP regions analysed with flare activity within 6 hours of a spike.}
    \begin{tabular}{cccccccc}
        \hline
        \multirow{2}{*}{Quantity} & Run. Mean & \multirow{2}{*}{$\sigma$} & Helicity Cutoff & \multirow{2}{*}{$S$} & \multirow{2}{*}{$S_w$} & \multirow{2}{*}{$\frac{1}{2} \left(S + S_W \right)$} & $N_s$ \\
        & (hr) & & (Mx$^2$) & & & & (non-flaring) \\
        \hline
        \dLc & 3 & 1.5 & 0 & 0.467 & 0.527 & 0.497 & 4752  \\
        \vzdl & 3 & 1.5 & 0 & 0.456 & 0.497 & 0.476 & 3712  \\
        \kurtvzdh & 3 & 1.5 & 0 & 0.447 & 0.5 & 0.473 & 7942  \\
        \dLc & 6 & 1.5 & 0 & 0.446 & 0.496 & 0.471 & 4499  \\
        \vzdh & 3 & 1.5 & 0 & 0.447 & 0.482 & 0.465 & 4192  \\
        \kurtvzbz & 3 & 1.5 & 0 & 0.427 & 0.494 & 0.461 & 7741  \\
        \kurtvzdl & 3 & 1.5 & 0 & 0.429 & 0.484 & 0.457 & 7796  \\
        \dLc & 3 & 2 & 0 & 0.428 & 0.485 & 0.456 & 2655  \\
        \dLc & 3 & 1.5 & 1$\times 10^{18}$ & 0.385 & 0.528 & 0.456 & 2672  \\
        \kurtdL & 3 & 1.5 & 0 & 0.417 & 0.487 & 0.452 & 7475  \\
        \hline
    \end{tabular}
    \label{tab:best6}
\end{table}
\begin{table}
    \centering
    \caption{10 best performing statistics in both $S$ and $S_w$ across all \FAR\ SHARP regions analysed with flare activity within 12 hours of a spike.}
    \begin{tabular}{cccccccc}
        \hline
        \multirow{2}{*}{Quantity} & Run. Mean & \multirow{2}{*}{$\sigma$} & Helicity Cutoff & \multirow{2}{*}{$S$} & \multirow{2}{*}{$S_w$} & \multirow{2}{*}{$\frac{1}{2} \left(S + S_W \right)$} & $N_s$ \\
        & (hr) & & (Mx$^2$) & & & & (non-flaring) \\
        \hline
        \dLc & 3 & 1.5 & 0 & 0.536 & 0.598 & 0.567 & 4752  \\
        \vzdl & 3 & 1.5 & 0 & 0.535 & 0.585 & 0.56 & 3712  \\
        \dLc & 6 & 1.5 & 0 & 0.529 & 0.587 & 0.558 & 4499  \\
        \kurtvzbz & 3 & 1.5 & 0 & 0.53 & 0.585 & 0.557 & 7741  \\
        \vzdh & 3 & 1.5 & 0 & 0.533 & 0.579 & 0.556 & 4192  \\
        \dLc & 3 & 2 & 0 & 0.519 & 0.593 & 0.556 & 2655  \\
        \vzbz & 3 & 1.5 & 0 & 0.529 & 0.582 & 0.556 & 4314  \\
        \kurtvzdh & 3 & 1.5 & 0 & 0.529 & 0.581 & 0.555 & 7942  \\
        \kurtvzbz & 3 & 2 & 0 & 0.524 & 0.58 & 0.552 & 4001  \\
        \kurtvzdh & 3 & 2 & 0 & 0.526 & 0.577 & 0.551 & 4117  \\
        \hline
    \end{tabular}
    \label{tab:best12}
\end{table}

\begin{figure*}
\centerline{\includegraphics[width=0.7\textwidth]{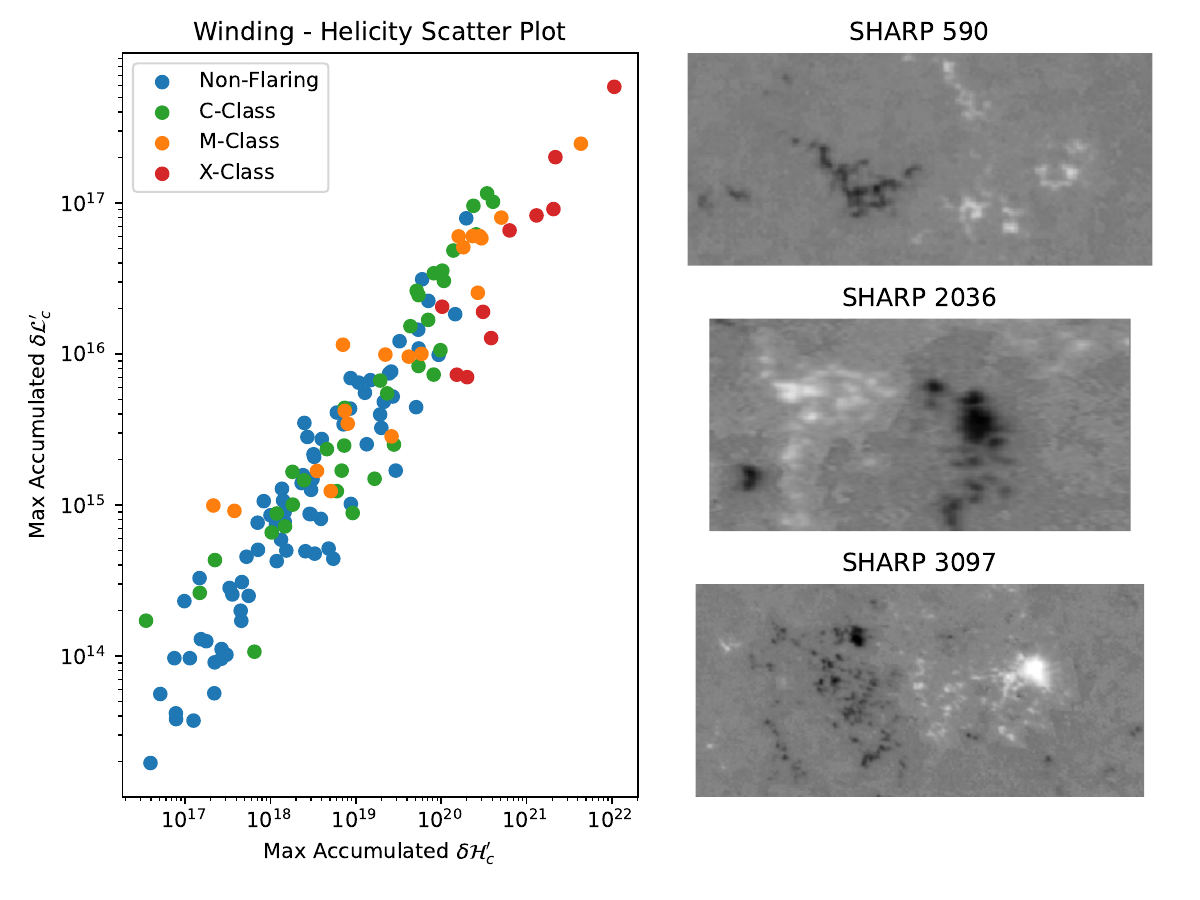}}
\caption{\textit{Left}: log-log scatter plot for maximum values of accumulated \dLc\ and \dHc\ for the \TF\ SHARP regions analysed. Colour is coded based on the class of the maximum flare magnitude given by the legend. \textit{Right}: three typical magnetograms from non-flaring SHARP regions.}
\label{fig:scatter}
\end{figure*}
The final point we make in this study is that a productive approach to the issue of the metrics being significantly misleading for non-flaring regions, might be to try to use the data to first classify regions as flaring or not. If one could identify the regions that do and do not flare, then the metrics could be applied to make individual flare predictions, and the above results indicate they have some potential in that regard, especially in the more active regions which have X-class flares.  In Figure\,\ref{fig:scatter}, we see a scatter plot for all \TF\ SHARP regions' maximum values for accumulated (time-integrated) \dLc\ and \dHc. Each region is colour-coded depending on the classification of the maximum flare magnitude associated with that region; X-class (\textit{red}), M-class (\textit{orange}), C-class (\textit{green}), and non-flaring (\textit{blue}). As one would expect, there is a strong correlation between the maximum accumulated \dLc\ and \dHc\ values given that magnetic winding is derived from magnetic helicity. However, for the flaring regions there does appear to be a positive trend in the amount of topology accumulated and the maximum magnitude of flare seen. More explicitly, all of the active regions with associated X-class flares are situated to the upper-right quadrant of the scatter plot, as are (proportionally) more of the active regions with M-class and C-class flares. If maximum accumulated values exceeding $6\times10^{15}\,\mathrm{cm}^{4}$ and $1\times10^{20}\,\mathrm{Mx}^2$ are seen for \dLc\ and \dHc, respectively, then there is a high probability of the region flaring (according to this dataset) with only three non-flaring regions exceeding these values from our dataset whilst all regions with X-class flares are captured. Additionally, visual inspection of the non-flaring region(s) in question, such as shown in the panels on the right of Figure\,\ref{fig:scatter}, reveals that many non-flaring regions are either small or in their decaying phases. As such, they exhibit diffuse magnetograms with no prominent concentrations of magnetic flux. This `diffusiveness', particularly for large SHARP patches, may explain how large accumulations of winding and helicity are not always associated with flares. That is, modest concentrations of topology are inputted over large areas. Subsequently, the localised concentrations required to create complex magnetic field configurations within the solar atmosphere that can become perturbed and trigger a reconnection event are not met and no flaring occurs. There are numerous existing methods for classifying regions \citep{mcintosh1990classification,kashapova2021analysis}, and in some cases automated \citep{de2020automatic}, which could be combined with these maximum \dLc\ and \dHc\ values to separate out flaring and non-flaring regions. We hope that by making this dataset available other researchers could effectively address this question.

In summary, the results presented in \S\,\ref{sec:results2} provide a valuable first step in transforming the ARTop topology calculations into predictive metrics for solar eruptions. Within flaring active regions, $\approx 40-60$\,\% of spikes and flares can be associated to one-another, with \dLc\ providing the most promising results/scores.

\section{Conclusions}\label{sec:conc}
This manuscript presents analysis of various quantities based of the topological fluxes calculated by the ARTop package, and assess their potential efficacy for flare prediction. They are applied to time-series of flaring across \TF\ (\FAR\ flaring; \NF\ non-flaring) active regions contained within SHARP datasets. The key findings from this study are listed as follows.

   \begin{enumerate}
      \item The $\delta$ measures of the winding and helicity rates provided by the ARTop package, which measure the net imbalance of current-carrying over potential field derived topological fluxes,  greatly reduces the dependence on the velocity smoothing ($VS$) parameter, which is used when creating velocity maps from magnetogram data using the DAVE4VM method. As this work highlights, more frequently employed topological quantities like the helicity and winding fluxes are sensitive to this choice, and depending on the values selected may mean photospheric signatures are not detected until after an eruption has occurred. 

      \item Constructing times-series (\dLc\ and \dHc) composed of only the current-carrying-dominant parts of the $\delta L$ and $\delta H$ fluxes produced time-series whose extremal values, spikes, (those outside of two standard deviations from the running mean over a suitable window) show significant temporal correlation with the timing of flares. When these signals are only classed as meaningful if a significant amount of net current-carrying helicity $\delta H$ has been accumulated in the region, this correlation improves significantly. It is further shown that a number of the spikes showed a spatio-temporal correlation with the sites of X-class flares which are co-temporal with CME events, providing evidence these spikes can provide physically meaningful signals.
       
      \item Across the parameter scans for the metrics investigated in this work, it seems that \dLc\ is the topological quantity with the greatest potential for forecasting flaring events on its own (based on spiking activity of the quantities examined during this study). Additionally, we found there is some benefit to adopting an helicity cutoff when determining if a spike should be counted as a meaningful event. We hypothesize that the cutoff acts as a means of quantifying whether there is sufficient complex magnetic field residing above the photosphere for new topology input to destabilise existing structures to cause reconnection events. Such an adoption yielded approximately half the number of false-positive spikes in non-flaring regions for \dLc\ with other, more stringent cutoffs, reducing this much further, albeit at the cost of further reduced efficacy for predicting flares. We present evidence that some algorithm which could deduce whether a region is likely to flare would allow the quantities derived in this study to form the basis for a predictive methodology for individual flare events.
      
      \item The majority of the parameter combinations presented in the data set (except \dLc) do not generate more spikes than there are flares for the subset of three regions (AR\,11158, 11302, 12673), which are analysed in detail. However, a similar spike-based analysis of the excess kurtosis for these metrics improved their ability to identify flares, becoming more comparable to \dLc, whilst also increasing the number of spikes such that there are more spikes than the number of flares. The excess kurtoses provide a compliment to the short-term (i.e. \dLc, \dHc, \vzbz, etc) and long-term (accumulated winding and helicity) metrics determined by ARTop, by providing a medium-term memory of the time-series. It is also shown in \S\,\ref{sec:kurtosis} that the kurtosis of time-series might be used in conjunction with the \dLc\ time-series to anticipate the magnitude of a flare (predictive signal).

   \end{enumerate}

The dataset used for this manuscript is publicly available\footnote{Github: Not available yet} as a pickle file along with some sample Jupyter Notebooks on how to process the data. This is a living database and will continually grow as more regions are processed with the ARTop code. This database will form the basis of a predictive model to further test the efficacy of topology calculations when it comes to forecasting the likelihood and potential magnitude of flaring events. As mentioned earlier, there are currently a large number of false positive spikes, especially in non-flaring regions, though it may be possible to segregate flaring and non-flaring regions based on the ratio of maximum accumulated winding and helicity and/or the visual appearance of magnetogram data, i.e. how diffuse/complex the region is. The issues identified here will be addressed in a following study.

\begin{acknowledgements}
The authors acknowledge support from the US Air Force grant FA8655-23-1-7247. D. M. acknowledges support from a Leverhulme Trust grant (RPG-2023-182), a Science and Technologies Facilities Council (STFC) grant (ST/Y001672/1) and a Personal Fellowship from the Royal Society of Edinburgh (ID:4282). C.P acknowledges support from The UK Science and Technology funding council under grant number ST/W00108X/1. This research utilises version 5.1.2 of the SunPy open source software package \citep{sunpypaper}.
\end{acknowledgements}

\bibliographystyle{aasjournal}
\bibliography{ref} 

\begin{thebibliography}{}
\expandafter\ifx\csname natexlab\endcsname\relax\def\natexlab#1{#1}\fi
\providecommand{\url}[1]{\href{#1}{#1}}
\providecommand{\dodoi}[1]{doi:~\href{http://doi.org/#1}{\nolinkurl{#1}}}
\providecommand{\doeprint}[1]{\href{http://ascl.net/#1}{\nolinkurl{http://ascl.net/#1}}}
\providecommand{\doarXiv}[1]{\href{https://arxiv.org/abs/#1}{\nolinkurl{https://arxiv.org/abs/#1}}}

\bibitem[{Alielden {et~al.}(2023)Alielden, MacTaggart, Ming, Prior, \&
  Raphaldini}]{artop1}
Alielden, K., MacTaggart, D., Ming, Q., Prior, C., \& Raphaldini, B. 2023, RAS
  Techniques and Instruments, 2, 398, \dodoi{10.1093/rasti/rzad029}

\bibitem[{{Aslam} {et~al.}(2024){Aslam}, {MacTaggart}, {Williams}, {Fletcher},
  \& {Romano}}]{mnraspaper}
{Aslam}, O., {MacTaggart}, D., {Williams}, T., {Fletcher}, L., \& {Romano}, P.
  2024, MNRAS, In Review

\bibitem[{Berger(1984)}]{berger1984rigorous}
Berger, M.~A. 1984, Geophysical \& Astrophysical Fluid Dynamics, 30, 79

\bibitem[{Bi {et~al.}(2018)Bi, Liu, Liu, Yang, Xu, \& Ji}]{bi2018survey}
Bi, Y., Liu, Y.~D., Liu, Y., {et~al.} 2018, The Astrophysical Journal, 865, 139

\bibitem[{Boers(2018)}]{boers2018early}
Boers, N. 2018, Nature communications, 9, 2556

\bibitem[{Chen {et~al.}(2012)Chen, Liu, Liu, Li, \& Aihara}]{chen2012detecting}
Chen, L., Liu, R., Liu, Z.-P., Li, M., \& Aihara, K. 2012, Scientific reports,
  2, 342

\bibitem[{Dakos {et~al.}(2019)Dakos, Matthews, Hendry, Levine, Loeuille,
  Norberg, Nosil, Scheffer, \& De~Meester}]{dakos2019ecosystem}
Dakos, V., Matthews, B., Hendry, A.~P., {et~al.} 2019, Nature ecology \&
  evolution, 3, 355

\bibitem[{de~Oliveira \& Gradvohl(2020)}]{de2020automatic}
de~Oliveira, L.~S., \& Gradvohl, A. L.~S. 2020, Revista Brasileira de
  Computa{\c{c}}{\~a}o Aplicada, 12, 67

\bibitem[{{Garland} {et~al.}(2024){Garland}, {Yurchyshyn}, {Loper}, {Akers}, \&
  {Emmons}}]{garland24}
{Garland}, S.~H., {Yurchyshyn}, V.~B., {Loper}, R.~D., {Akers}, B.~F., \&
  {Emmons}, D.~J. 2024, Frontiers in Astronomy and Space Sciences, In Press

\bibitem[{{Georgoulis} {et~al.}(2021){Georgoulis}, {Bloomfield}, {Piana},
  {Massone}, {Soldati}, {Gallagher}, {Pariat}, {Vilmer}, {Buchlin}, {Baudin},
  {Csillaghy}, {Sathiapal}, {Jackson}, {Alingery}, {Benvenuto}, {Campi},
  {Florios}, {Gontikakis}, {Guennou}, {Guerra}, {Kontogiannis}, {Latorre},
  {Murray}, {Park}, {von Stachelski}, {Torbica}, {Vischi}, \&
  {Worsfold}}]{2021JSWSC}
{Georgoulis}, M.~K., {Bloomfield}, D.~S., {Piana}, M., {et~al.} 2021, Journal
  of Space Weather and Space Climate, 11, 39, \dodoi{10.1051/swsc/2021023}

\bibitem[{{Hoeksema} {et~al.}(2014){Hoeksema}, {Liu}, {Hayashi}, {Sun},
  {Schou}, {Couvidat}, {Norton}, {Bobra}, {Centeno}, {Leka}, {Barnes}, \&
  {Turmon}}]{hoeksema14}
{Hoeksema}, J.~T., {Liu}, Y., {Hayashi}, K., {et~al.} 2014, \solphys, 289,
  3483, \dodoi{10.1007/s11207-014-0516-8}

\bibitem[{{Hurlburt}(2022)}]{hekpaper}
{Hurlburt}, N.~E. 2022, in AGU Fall Meeting Abstracts, Vol. 2022, SH32F--1818

\bibitem[{Jarolim {et~al.}(2023)Jarolim, Thalmann, Veronig, \&
  Podladchikova}]{jarolim2023probing}
Jarolim, R., Thalmann, J., Veronig, A., \& Podladchikova, T. 2023, Nature
  Astronomy, 7, 1171

\bibitem[{Kashapova {et~al.}(2021)Kashapova, Zhukova, Miteva, Zhdanov,
  Myagkova, \& Meshalkina}]{kashapova2021analysis}
Kashapova, L., Zhukova, A., Miteva, R., {et~al.} 2021, Geomagnetism and
  Aeronomy, 61, 1022

\bibitem[{Knizhnik {et~al.}(2015)Knizhnik, Antiochos, \&
  DeVore}]{knizhnik2015filament}
Knizhnik, K.~J., Antiochos, S.~K., \& DeVore, C.~R. 2015, The Astrophysical
  Journal, 809, 137

\bibitem[{Kors{\'o}s {et~al.}(2022)Kors{\'o}s, Erd{\'e}lyi, Huang, \&
  Morgan}]{korsos2022magnetic}
Kors{\'o}s, M., Erd{\'e}lyi, R., Huang, X., \& Morgan, H. 2022, The
  Astrophysical Journal, 933, 66

\bibitem[{Kors{\'o}s {et~al.}(2020)Kors{\'o}s, Romano, Morgan, Ye, Erd{\'e}lyi,
  \& Zuccarello}]{korsos2020differences}
Kors{\'o}s, M., Romano, P., Morgan, H., {et~al.} 2020, The Astrophysical
  Journal Letters, 897, L23

\bibitem[{Kusano {et~al.}(2020)Kusano, Iju, Bamba, \&
  Inoue}]{kusano2020physics}
Kusano, K., Iju, T., Bamba, Y., \& Inoue, S. 2020, Science, 369, 587

\bibitem[{LaBonte {et~al.}(2007)LaBonte, Georgoulis, \&
  Rust}]{labonte2007survey}
LaBonte, B., Georgoulis, M., \& Rust, D. 2007, The Astrophysical Journal, 671,
  955

\bibitem[{Leka {et~al.}(2019{\natexlab{a}})Leka, Park, Kusano, Andries, Barnes,
  Bingham, Bloomfield, McCloskey, Delouille, Falconer,
  {et~al.}}]{leka2019comparison}
Leka, K., Park, S.-H., Kusano, K., {et~al.} 2019{\natexlab{a}}, The
  Astrophysical Journal Supplement Series, 243, 36

\bibitem[{Leka {et~al.}(2019{\natexlab{b}})Leka, Park, Kusano, Andries, Barnes,
  Bingham, Bloomfield, McCloskey, Delouille, Falconer,
  {et~al.}}]{leka2019comparison3}
---. 2019{\natexlab{b}}, The Astrophysical Journal, 881, 101

\bibitem[{{Lemen} {et~al.}(2012){Lemen}, {Title}, {Akin}, {Boerner}, {Chou},
  {Drake}, {Duncan}, {Edwards}, {Friedlaender}, {Heyman}, {Hurlburt}, {Katz},
  {Kushner}, {Levay}, {Lindgren}, {Mathur}, {McFeaters}, {Mitchell}, {Rehse},
  {Schrijver}, {Springer}, {Stern}, {Tarbell}, {Wuelser}, {Wolfson}, {Yanari},
  {Bookbinder}, {Cheimets}, {Caldwell}, {Deluca}, {Gates}, {Golub}, {Park},
  {Podgorski}, {Bush}, {Scherrer}, {Gummin}, {Smith}, {Auker}, {Jerram},
  {Pool}, {Soufli}, {Windt}, {Beardsley}, {Clapp}, {Lang}, \&
  {Waltham}}]{lemen12}
{Lemen}, J.~R., {Title}, A.~M., {Akin}, D.~J., {et~al.} 2012, Solar Physics,
  275, 17, \dodoi{10.1007/s11207-011-9776-8}

\bibitem[{Liu {et~al.}(2019)Liu, Cheng, Wang, \& Zhou}]{liu2019formation}
Liu, L., Cheng, X., Wang, Y., \& Zhou, Z. 2019, The Astrophysical Journal, 884,
  45

\bibitem[{Liu \& Schuck(2012)}]{liu2012magnetic}
Liu, Y., \& Schuck, P. 2012, The Astrophysical Journal, 761, 105

\bibitem[{Liu {et~al.}(2023)Liu, Welsch, Valori, Georgoulis, Guo, Pariat, Park,
  \& Thalmann}]{liu2023changes}
Liu, Y., Welsch, B.~T., Valori, G., {et~al.} 2023, The Astrophysical Journal,
  942, 27

\bibitem[{Lumme {et~al.}(2019)Lumme, Kazachenko, Fisher, Welsch, Pomoell, \&
  Kilpua}]{lumme2019probing}
Lumme, E., Kazachenko, M., Fisher, G., {et~al.} 2019, Solar Physics, 294, 84

\bibitem[{{MacTaggart} {et~al.}(2021){MacTaggart}, {Prior}, {Raphaldini},
  {Romano}, \& {Guglielmino}}]{mactaggart21}
{MacTaggart}, D., {Prior}, C., {Raphaldini}, B., {Romano}, P., \&
  {Guglielmino}, S.~L. 2021, Nature Communications, 12, 6621,
  \dodoi{10.1038/s41467-021-26981-7}

\bibitem[{McIntosh(1990)}]{mcintosh1990classification}
McIntosh, P.~S. 1990, Solar Physics, 125, 251

\bibitem[{Moraitis {et~al.}(2024)Moraitis, Patsourakos, Nindos, Thalmann, \&
  Pariat}]{moraitis2024using}
Moraitis, K., Patsourakos, S., Nindos, A., Thalmann, J., \& Pariat, {\'E}.
  2024, Astronomy \& Astrophysics, 683, A87

\bibitem[{Moraitis {et~al.}(2019)Moraitis, Sun, Pariat, \&
  Linan}]{moraitis2019magnetic}
Moraitis, K., Sun, X., Pariat, {\'E}., \& Linan, L. 2019, Astronomy \&
  Astrophysics, 628, A50

\bibitem[{{Mumford} {et~al.}(2024){Mumford}, {Freij}, {Stansby}, {Christe},
  {Ireland}, {Shih}, {Mayer}, {Hughitt}, {Ryan}, {Liedtke}, {Hayes}, {Barnes},
  {P{\'e}rez-Su{\'a}rez}, {I.}, {Chakraborty}, {Inglis}, {Pattnaik},
  {Sip{\H{o}}cz}, {MacBride}, {Sharma}, {Leonard}, {Hewett}, {Hamilton},
  {Manhas}, {Panda}, {Earnshaw}, {Choudhary}, {Kumar}, {Singh}, {Chanda},
  {Akramul Haque}, {Kirk}, {Mueller}, {Konge}, {Wentzel-Long}, {Srivastava},
  {Maloney}, {Jain}, {Zivadinovic}, {Bennett}, {Wilson}, {Baruah}, {Arbolante},
  {Simon}, {Charlton}, {Mishra}, {Paul}, {Verma}, {Chorley}, {Chouhan},
  {Himanshu}, {Mason}, {Modi}, {Sharma}, {Naman9639}, {Bobra}, {Tyagi}, {Campos
  Rozo}, {Manley}, {Ivashkiv}, {Laitinen}, {Chatterjee}, {von Forstner},
  {Baz{\'a}n}, {Akira Stern}, {Shukla}, {Gieseler}, {Evans}, {Jain}, {Malocha},
  {Ghosh}, {Airmansmith97}, {Sta{\'n}czak}, {Ranjan Singh}, {De Visscher},
  {Verma}, {SophieLemos}, {Agrawal}, {Alam}, {Buddhika}, {Collier}, {Pathak},
  {Rideout}, {Sharma}, {Park}, {Bates}, {ryuusama}, {Shukla}, {Giger},
  {Mishra}, {Sharma}, {Goel}, {Taylor}, {Cetusic}, {Reiter}, {Jacob},
  {Inchaurrandieta}, {Dacie}, {Dubey}, {Eigenbrot}, {Bray}, {Murphy}, {Surve},
  {Zahniy}, {Sidhu}, {Meszaros}, {Parkhi}, {Russell}, {Bose}, {Pandey},
  {Price-Whelan}, {J}, {Chicrala}, {Ankit}, {Graham}, {Guennou}, {D'Avella},
  {Williams}, {Verma}, {Ballew}, {Agrawal}, {Singh}, {Lodha}, {Robitaille},
  {Augspurger}, {Krishan}, {honey}, {neerajkulk}, {Bhope}, {Singh Gaba},
  {Hill}, {Dixit}, {Mampaey}, {Wiedemann}, {Molina}, {Garcia Briseno},
  {Ke{\c{s}}kek}, {Habib}, {Letts}, {Singaravelan}, {Ranjan}, {Altunian},
  {Streicher}, {Gomillion}, {Agarwal}, {Kothari}, {Nomiya}, {mridulpandey},
  {Stevens}, {B}, {Bahuleyan}, {Shauryam}, {Kaszynski}, {W}, {Mehrotra},
  {Tang}, {Sinha}, {Smith}, {Sinha}, {Kustov}, {Stone}, {Bard}, {Behn},
  {Arias}, {Tollerud}, {Mackenzie Dover}, {Verstringe}, {Kumar}, {Mathur},
  {Babuschkin}, {Calixto}, {Wimbish}, {Qing}, {Buitrago-Casas}, {Krishna},
  {Chaudhari}, {Hiware}, {Ghosh}, {McKee}, {Lyes}, {Mangaonkar}, {Cheung},
  {Mendero}, {Dedhia}, {Schoentgen}, {Shahdadpuri}, {Srinivasan}, {Gyenge},
  {OussCHE}, {Wright}, {Reddy Mekala}, {Das}, {Mishra}, {Sharma}, {Badman},
  {Van Kooten}, {Srikanth}, {Jain}, {Farah}, {Kannojia}, {Mihan Chistie},
  {Qing}, {Yadav}, {Paul}, {Wilkinson}, {Caswell}, {Braccia}, {Pereira},
  {Gates}, {Kien Dang}, {Bankar}, {Jamieson}, {Agrawal}, {ejm4567}, {platipo},
  {pradeep}, {resakra}, {tal66}, {yasintoda}, {Attie}, \&
  {Murray}}]{sunpypaper}
{Mumford}, S.~J., {Freij}, N., {Stansby}, D., {et~al.} 2024, {SunPy}, v5.1.1,
  Zenodo, \dodoi{10.5281/zenodo.591887}

\bibitem[{Nindos {et~al.}(2003)Nindos, Zhang, \& Zhang}]{nindos2003magnetic}
Nindos, A., Zhang, J., \& Zhang, H. 2003, The Astrophysical Journal, 594, 1033

\bibitem[{Pariat {et~al.}(2005)Pariat, D{\'e}moulin, \&
  Berger}]{pariat2005photospheric}
Pariat, E., D{\'e}moulin, P., \& Berger, M.~A. 2005, Astronomy \& Astrophysics,
  439, 1191

\bibitem[{Pariat {et~al.}(2017)Pariat, Leake, Valori, Linton, Zuccarello, \&
  Dalmasse}]{pariat2017relative}
Pariat, E., Leake, J., Valori, G., {et~al.} 2017, Astronomy \& Astrophysics,
  601, A125

\bibitem[{Pariat {et~al.}(2006)Pariat, Nindos, D{\'e}moulin, \&
  Berger}]{pariat2006spatial}
Pariat, E., Nindos, A., D{\'e}moulin, P., \& Berger, M. 2006, Astronomy \&
  Astrophysics, 452, 623

\bibitem[{Park {et~al.}(2010)Park, Chae, \& Wang}]{park2010productivity}
Park, S.-h., Chae, J., \& Wang, H. 2010, The Astrophysical Journal, 718, 43

\bibitem[{Park {et~al.}(2008)Park, Lee, Choe, Chae, Jeong, Yang, Jing, \&
  Wang}]{park2008variation}
Park, S.-H., Lee, J., Choe, G.-S., {et~al.} 2008, The Astrophysical Journal,
  686, 1397

\bibitem[{Park {et~al.}(2021)Park, Leka, \& Kusano}]{park2021magnetic}
Park, S.-H., Leka, K., \& Kusano, K. 2021, The Astrophysical Journal, 911, 79

\bibitem[{Pevtsov {et~al.}(2003)Pevtsov, Maleev, \&
  Longcope}]{pevtsov2003helicity}
Pevtsov, A.~A., Maleev, V.~M., \& Longcope, D.~W. 2003, The Astrophysical
  Journal, 593, 1217

\bibitem[{{Price} {et~al.}(2019){Price}, {Pomoell}, {Lumme}, \&
  {Kilpua}}]{price19}
{Price}, D.~J., {Pomoell}, J., {Lumme}, E., \& {Kilpua}, E.~K.~J. 2019, \aap,
  628, A114, \dodoi{10.1051/0004-6361/201935535}

\bibitem[{Priest {et~al.}(2016)Priest, Longcope, \&
  Janvier}]{priest2016evolution}
Priest, E.~R., Longcope, D., \& Janvier, M. 2016, Solar Physics, 291, 2017

\bibitem[{Prior \& MacTaggart(2020)}]{prior2020magnetic}
Prior, C., \& MacTaggart, D. 2020, Proceedings of the Royal Society A, 476,
  20200483

\bibitem[{Raphaldini {et~al.}(2023)Raphaldini, Dikpati, Norton, Teruya,
  McIntosh, Prior, \& MacTaggart}]{raphaldini2023deciphering}
Raphaldini, B., Dikpati, M., Norton, A.~A., {et~al.} 2023, The Astrophysical
  Journal, 958, 175

\bibitem[{{Raphaldini} {et~al.}(2022){Raphaldini}, {Prior}, \&
  {MacTaggart}}]{brenopaper}
{Raphaldini}, B., {Prior}, C.~B., \& {MacTaggart}, D. 2022, \apj, 927, 156,
  \dodoi{10.3847/1538-4357/ac4df9}

\bibitem[{{Schou} {et~al.}(2012){Schou}, {Scherrer}, {Bush}, {Wachter},
  {Couvidat}, {Rabello-Soares}, {Bogart}, {Hoeksema}, {Liu}, {Duvall}, {Akin},
  {Allard}, {Miles}, {Rairden}, {Shine}, {Tarbell}, {Title}, {Wolfson},
  {Elmore}, {Norton}, \& {Tomczyk}}]{hmipaper}
{Schou}, J., {Scherrer}, P.~H., {Bush}, R.~I., {et~al.} 2012, \solphys, 275,
  229, \dodoi{10.1007/s11207-011-9842-2}

\bibitem[{Schuck(2008)}]{schuck08}
Schuck, P.~W. 2008, The Astrophysical Journal, 683, 1134,
  \dodoi{10.1086/589434}

\bibitem[{Sevim {et~al.}(2014)Sevim, Oztekin, Bali, Gumus, \&
  Guresen}]{sevim2014developing}
Sevim, C., Oztekin, A., Bali, O., Gumus, S., \& Guresen, E. 2014, European
  Journal of Operational Research, 237, 1095

\bibitem[{Thalmann {et~al.}(2019)Thalmann, Moraitis, Linan, Pariat, Valori, \&
  Dalmasse}]{thalmann2019magnetic}
Thalmann, J.~K., Moraitis, K., Linan, L., {et~al.} 2019, The Astrophysical
  Journal, 887, 64

\bibitem[{Tziotziou {et~al.}(2013)Tziotziou, Georgoulis, \&
  Liu}]{tziotziou2013interpreting}
Tziotziou, K., Georgoulis, M.~K., \& Liu, Y. 2013, The Astrophysical Journal,
  772, 115

\bibitem[{Vemareddy(2019)}]{vemareddy2019very}
Vemareddy, P. 2019, The Astrophysical Journal, 872, 182

\bibitem[{Vemareddy(2021)}]{vemareddy2021successive}
---. 2021, Monthly Notices of the Royal Astronomical Society, 507, 6037

\bibitem[{Wang {et~al.}(2018)Wang, Liu, Hoeksema, Zimovets, \&
  Liu}]{wang2018roles}
Wang, R., Liu, Y.~D., Hoeksema, J.~T., Zimovets, I., \& Liu, Y. 2018, The
  Astrophysical Journal, 869, 90

\bibitem[{{Williams} \& {Morgan}(2022)}]{almanac}
{Williams}, T., \& {Morgan}, H. 2022, Space Weather, 20, e2022SW003253,
  \dodoi{10.1029/2022SW003253}

\bibitem[{Zuccarello {et~al.}(2018)Zuccarello, Pariat, Valori, \&
  Linan}]{zuccarello2018threshold}
Zuccarello, F.~P., Pariat, E., Valori, G., \& Linan, L. 2018, The Astrophysical
  Journal, 863, 41

\end{thebibliography}

\begin{appendix}      
Here, the full list of \TF\ SHARP regions used for this study are given in Table\,\ref{tab:sharps_appendix}. A breakdown on the largest flare size, as well as the total number of X-, M-, and C-class flares are given for each region. In total, \FAR\ regions exhibit flaring, with a further \NF\ regions being flare free.

\begin{center}
\begin{longtable}{ c c c c c c }
\caption{SHARP regions investigated with flare information provided by \textit{HEK}.}\label{tab:sharps_appendix} \\

\hline \multicolumn{1}{c}{NOAA Active Region} & \multicolumn{1}{c}{SHARP Number} & \multicolumn{1}{c}{Largest Flare} & \multicolumn{1}{c}{\# X--class Flares} & \multicolumn{1}{c}{\# M--class Flares} & \multicolumn{1}{c}{\# C--class Flares} \\ 
\endfirsthead
\hline                   

11069  &  8  &  M1.2  &  ---  &  1  &  7  \\
11070  &  14  & --- & --- & --- & --- \\
11071  &  17  & --- & --- & --- & --- \\
11072  &  26  & --- & --- & --- & --- \\
11076  &  43  & --- & --- & --- & --- \\
11073/77  &  45  & --- & --- & --- & --- \\
11079  &  49  &  M1.0  &  ---  &  1  &  ---  \\
11080  &  51  &  C1.2  &  ---  &  ---  &  1  \\
11081  &  54  &  M2.0  &  ---  &  1  &  5  \\
11086  &  83  & --- & --- & --- & --- \\
11087  &  86  &  C3.4  &  ---  &  ---  &  5  \\
11096  &  116  & --- & --- & --- & --- \\
11098  &  131  & --- & --- & --- & --- \\
11105  &  156  &  C3.3  &  ---  &  ---  &  2  \\
11114  &  219  & --- & --- & --- & --- \\
11116  &  221  & --- & --- & --- & --- \\
11130  &  274  & --- & --- & --- & --- \\
11132  &  285  & --- & --- & --- & --- \\
11136  &  316  & --- & --- & --- & --- \\
11138  &  318  &  C1.3  &  ---  &  ---  &  1  \\
11141  &  325  &  C1.9  &  ---  &  ---  &  1  \\
11143  &  335  & --- & --- & --- & --- \\
11148  &  347  & --- & --- & --- & --- \\
11151  &  354  & --- & --- & --- & --- \\
11155  &  366  & --- & --- & --- & --- \\
11156  &  367  & --- & --- & --- & --- \\
11158  &  377  &  X2.2  &  1  &  5  &  54  \\
11160/61/62  &  384  &  M1.3  &  ---  &  4  &  20  \\
11165  &  394  &  M5.3  &  ---  &  6  &  24  \\
11172/75  &  421  &  C1.6  &  ---  &  ---  &  2  \\
11173  &  429  & --- & --- & --- & --- \\
11179  &  436  & --- & --- & --- & --- \\
11176/78  &  437  &  M1.4  &  ---  &  3  &  13  \\
11177  &  438  & --- & --- & --- & --- \\
11198  &  527  & --- & --- & --- & --- \\
11199  &  540  &  C6.5  &  ---  &  ---  &  5  \\
11206  &  572  & --- & --- & --- & --- \\
11209  &  589  & --- & --- & --- & --- \\
11212  &  595  & --- & --- & --- & --- \\
11214/17  &  602  & --- & --- & --- & --- \\
11221  &  619  & --- & --- & --- & --- \\
11219/24  &  622  &  C5.9  &  ---  &  ---  &  2  \\
11223  &  625  &  C1.4  &  ---  &  ---  &  2  \\
11242  &  686  & --- & --- & --- & --- \\
11245/53  &  700  & --- & --- & --- & --- \\
11248/53/57  &  705  & --- & --- & --- & --- \\
11258  &  713  & --- & --- & --- & --- \\
11273  &  799  & --- & --- & --- & --- \\
11281  &  824  & --- & --- & --- & --- \\
11283  &  833  &  X2.1  &  2  &  5  &  13  \\
11291  &  851  & --- & --- & --- & --- \\
11300  &  875  & --- & --- & --- & --- \\
---  &  877  &  C9.6  &  ---  &  ---  &  15  \\
11302  &  892  &  X1.9  &  2  &  15  &  31  \\
11311  &  926  & --- & --- & --- & --- \\
11314/19  &  940  &  M1.3  &  ---  &  2  &  32  \\
11318  &  956  & --- & --- & --- & --- \\
11327  &  982  & --- & --- & --- & --- \\
11326  &  990  & --- & --- & --- & --- \\
11339/48  &  1028  &  X1.9  &  1  &  11  &  47  \\
11341/42  &  1041  &  M1.1  &  ---  &  1  &  3  \\
11357  &  1080  &  C1.8  &  ---  &  ---  &  3  \\
11373  &  1170  & --- & --- & --- & --- \\
11380/87  &  1209  &  M4.0  &  ---  &  3  &  11  \\
11385  &  1232  & --- & --- & --- & --- \\
11398  &  1303  & --- & --- & --- & --- \\
11397  &  1312  & --- & --- & --- & --- \\
11420  &  1399  & --- & --- & --- & --- \\
11429/30  &  1449  &  X5.4  &  3  &  14  &  36  \\
11434  &  1464  & --- & --- & --- & --- \\
11450  &  1528  &  C3.1  &  ---  &  ---  &  5  \\
11463  &  1558  &  C8.9  &  ---  &  ---  &  5  \\
11464  &  1594  & --- & --- & --- & --- \\
11469/73  &  1611  &  C2.6  &  ---  &  ---  &  12  \\
11477/78  &  1644  & --- & --- & --- & --- \\
11488  &  1688  & --- & --- & --- & --- \\
11527/28  &  1877  &  C5.0  &  ---  &  ---  &  7  \\
11531  &  1886  &  C1.7  &  ---  &  ---  &  2  \\
11546  &  1942  & --- & --- & --- & --- \\
11562  &  1990  &  C8.4  &  ---  &  ---  &  2  \\
11561  &  1997  & --- & --- & --- & --- \\
11565  &  2007  &  C2.3  &  ---  &  ---  &  2  \\
11572  &  2036  & --- & --- & --- & --- \\
11591  &  2121  & --- & --- & --- & --- \\
11601  &  2158  & --- & --- & --- & --- \\
11613/17  &  2191  &  M6.0  &  ---  &  5  &  15  \\
11628/29  &  2262  &  C5.7  &  ---  &  ---  &  3  \\
11651  &  2348  & --- & --- & --- & --- \\
11668  &  2436  & --- & --- & --- & --- \\
11682  &  2501  & --- & --- & --- & --- \\
11696  &  2560  &  C2.2  &  ---  &  ---  &  2  \\
11719  &  2635  &  M6.5  &  ---  &  2  &  13  \\
11737  &  2711  & --- & --- & --- & --- \\
11752  &  2754  &  C1.3  &  ---  &  ---  &  3  \\
11768  &  2832  & --- & --- & --- & --- \\
11784  &  2922  &  C1.1  &  ---  &  ---  &  1  \\
11796  &  2976  & --- & --- & --- & --- \\
11809  &  3028  &  C4.9  &  ---  &  ---  &  9  \\
11824  &  3097  & --- & --- & --- & --- \\
11835  &  3122  & --- & --- & --- & --- \\
11881  &  3309  & --- & --- & --- & --- \\
11892  &  3353  & --- & --- & --- & --- \\
11905  &  3420  &  C3.3  &  ---  &  ---  &  6  \\
11916  &  3448  &  C3.3  &  ---  &  ---  &  6  \\
11930  &  3515  &  C1.8  &  ---  &  ---  &  3  \\
11942  &  3560  & --- & --- & --- & --- \\
11967/72/75  &  3686  &  M6.6  &  ---  &  23  &  66  \\
11978  &  3741  & --- & --- & --- & --- \\
11996  &  3813  &  M9.3  &  ---  &  4  &  10  \\
12009  &  3853  & --- & --- & --- & --- \\
12024  &  3907  &  C2.4  &  ---  &  ---  &  1  \\
12036/37/43  &  3999  &  M7.3  &  ---  &  1  &  27  \\
12050  &  4075  &  C1.1  &  ---  &  ---  &  1  \\
12066  &  4131  & --- & --- & --- & --- \\
12089  &  4231  & --- & --- & --- & --- \\
12111  &  4328  & --- & --- & --- & --- \\
12192  &  4698  &  X3.1  &  6  &  32  &  72  \\
12268/70/79  &  5107  &  M2.1  &  ---  &  6  &  31  \\
12297  &  5298  &  X2.1  &  1  &  22  &  92  \\
12673  &  7115  &  X9.3  &  4  &  26  &  52  \\
11211  &  590  & --- & --- & --- & --- \\
11121/23  &  245  &  M5.4  &  ---  &  3  &  18  \\
11416  &  1389  &  C1.5  &  ---  &  ---  &  1  \\
11437  &  1480  & --- & --- & --- & --- \\
11446  &  1497  & --- & --- & --- & --- \\
11456  &  1564  & --- & --- & --- & --- \\
11460/64  &  1578  &  C3.7  &  ---  &  ---  &  5  \\
11523  &  1863  & --- & --- & --- & --- \\
11547  &  1943  & --- & --- & --- & --- \\
11549  &  1948  & --- & --- & --- & --- \\
11476  &  1638  &  M5.7  &  ---  &  11  &  84  \\
11821  &  3079  & --- & --- & --- & --- \\
11748  &  2748  &  X3.2  &  3  &  3  &  18  \\
11640  &  2337  &  C4.0  &  ---  &  ---  &  6  \\
11664  &  2425  & --- & --- & --- & --- \\
11598  &  2137  &  X1.8  &  1  &  3  &  23  \\
11560  &  1993  &  M1.6  &  ---  &  1  &  17  \\
11466/68  &  1603  &  M1.0  &  ---  &  1  &  4  \\
11680  &  2504  & --- & --- & --- & --- \\
11631/32  &  2291  &  C1.4  &  ---  &  ---  &  6  \\
11630  &  2270  &  C5.5  &  ---  &  ---  &  2  \\
11568  &  2017  &  C1.7  &  ---  &  ---  &  1  \\
11548  &  1946  &  M5.5  &  ---  &  5  &  11  \\
11554  &  1962  &  C7.6  &  ---  &  ---  &  5  \\

\hline

\end{longtable}
\end{center}

\end{appendix}

\end{document}